\documentclass[10pt,preprint]{aastex}

\def\Cosmospp{{\textit{Cosmos{\small++}}}\ }

\slugcomment{To be Submitted to ApJ}

\begin{document}

\title{ Numerical Modeling of the Radio Nebula from the 2004 December 27 Giant Flare of SGR 1806-20 }

\author{Jay D. Salmonson\altaffilmark{1,2},
P. Chris Fragile\altaffilmark{3}, and Peter Anninos\altaffilmark{1}}
\altaffiltext{1}{University of California, Lawrence Livermore
National Laboratory, Livermore CA 94550 }
\altaffiltext{2}{salmonson@llnl.gov} \altaffiltext{3}{College of
Charleston, Charleston, SC 29424; fragilep@cofc.edu}

\date{{\small    \today}}
\date{{\small   \LaTeX-ed \today}}

\begin{abstract}
We use the relativistic hydrodynamics code \Cosmospp to model the
evolution of the radio nebula triggered by the Dec.~27, 2004 giant
flare event of soft gamma repeater 1806-20. We primarily focus on
the rebrightening and centroid motion occurring subsequent to day 20
following the flare event. We model this period as a mildly
relativistic ($\gamma \sim 1.07 - 1.67$) jetted outflow expanding
into the interstellar medium (ISM). We demonstrate that a jet with
total energy $\sim 10^{46}$ ergs confined to a half opening angle
$\sim 20^\circ$ fits the key observables of this event, e.g.~the
flux lightcurve, emission map centroid position, and aspect ratio.
In particular, we find excellent agreement with observations if the
rebrightening is due to the jet, moving at $0.5 c$ and inclined
$\sim 0^\circ - 40^\circ$ toward the observer, colliding with a
density discontinuity in the ISM at a radius of several $10^{16}$ cm.
We also find that a jet with a higher velocity, $ \gtrsim 0.7 c$,
and larger inclination, $\gtrsim 70^\circ$, moving into a uniform
ISM can fit the observations in general, but tends to miss the
details of rebrightening.  The latter, uniform ISM model predicts an
ISM density more than 100 times lower than that of the former model,
and thus suggests an independent test which might discriminate
between the two.  One of the strongest constraints of both models is
that the data seems to require a non-uniform jet in order to be well
fit.
\end{abstract}


\section{Introduction}
\label{sec:intro}

On Dec.~27, 2004 a giant flare (GF) of $\gamma$-rays was observed
from soft gamma repeater (SGR) 1806-20.  This GF, the largest
$\gamma$-ray transient event ever observed from within our galaxy,
was comprised of a 0.2 s burst so bright that it briefly saturated
all $\gamma$ and X-ray detectors in the solar system, followed by a
several 100 s fading X-ray transient \citep{hbs+05}.  Such an
outburst is thought to derive from a cataclysmic reorganization of
the surface magnetic field of the magnetar, a neutron star
distinguished by its exceptionally high surface magnetic field,
$\sim 10^{15}$ G \citep{dt92}, which is the source of the SGR.

What is more, about one week after the GF a radio nebula was observed
at the location of the SGR \citep{gkgt+05}.  Over the next couple of
months this radio nebula was seen to expand and move across the sky
\citep{tggg+05,gleg+05}.  The observations suggest two epochs of
evolution.  Epoch I spanned from the first detection of this radio
afterglow about 7 days after the flare event to about 20 days post
flare. This epoch was distinguished by a steep decay in observed flux
($\sim t^{-2.7}$), a slight motion of the flux centroid toward the
West, and the detection of polarization of a few percent
\citep{gkgt+05}.  Also, the nebula was observed to be elongated along
the direction of motion with a roughly 2:1 aspect ratio.  As pointed
out by several authors, this epoch had interesting substructure,
including a steepening of, or break in, the flux decay rate at $\sim
9$ days as well as variation in amplitude and pitch angle of observed
polarization \citep{gleg+05,tggg+05,ccrk+05}. The observation of
polarization is indicative of some global order to the magnetic field
hosting the synchrotron electrons.

Epoch II, commencing at about 20 days, was distinct from Epoch I in
several ways.  First, a rapid rebrightening occurred, followed by a
shallower flux decay law ($\sim t^{-1.1}$) that persisted for the
remainder of the reported observations, until $\sim 80$ days.
Secondly the centroid of the flux began to move rapidly toward the
Northwest. Also, polarization abruptly vanished, with no significant
detections during this entire period, indicating a separate and
distinct population of radiating electrons from that of Epoch I
\citep{gleg+05,tggg+05}.

The detection of proper motion of the emission source strongly
suggests some sort of asymmetry.  This asymmetry could be rooted in
{\it i)} the ejected material having a particular direction, or {\it
ii)} the surrounding environment having a particular shape.  While
nature might easily incorporate both components, the models proposed
thus far, including the bulk of our work here, fall into the former
category. We have, however, briefly explored the latter possibility
by simulating a spherical isotropic mass ejection event centered in
a cylindrical, evacuated cavity.  This geometry mocks up the cavity
that could be swept out by the SGR's spin-down luminosity as it
moves through the interstellar medium (ISM). In this model, the
source of asymmetry would be the unhindered propagation of ejecta
down the cylinder, in contrast to the ejecta colliding with its
wall. Instead, the result we found was that the ejecta progressing
down the cylinder, diminishing as $1/r^2$, is quickly dwarfed by the
nearly isotropic external shock.  Thus any early asymmetry is washed
out.  As such, this model makes the strong prediction that the flux
centroid should return to the position of the SGR after some time.
Since this has not been observed to happen for SGR 1806-20, such a
model is disfavored.

The rest of the paper is organized into four sections.  We begin by
describing the models we wish to investigate and frame the parameter
space to be explored with some theory, \S\ref{sec:models}. In
particular, we study two distinct models for the emission of Epoch II:
collisional brightening, \S\ref{sec:collision}, in which this epoch
commences when ejecta collides and shocks with a density step in the
ISM, and doppler brightening, \S\ref{sec:doppler}, for which the
ejecta's motion through a uniform ISM, and its progressive shocking
and decelaration, is responsible for the Epoch II rebrightening.  The
latter model is most similar to that put forward by \citet{grrte+06},
except that we assume Epoch I is independent of Epoch II (see
\S\ref{sec:models}).  We also investigate structured jets
\S\ref{sec:structjets} which we find to play a surprisingly important
and interesting role in fitting the data.  We then describe our
numerical modeling method, \S\ref{sec:modeling}, including our novel
approach to post-processing observations from the data,
\S\ref{sec:postprocessing}.  We describe our results in
\S\ref{sec:results}.  While we find that both models studied are
viable, the collisional brightening model,
\S\ref{sec:collisionalresults}, does a better job than the doppler
brightening model, \S\ref{sec:dopplerresults}, of fitting the data,
particularly that of the lightcurve and centroid position.  Of
particular note, we find that structured jets,
\S\ref{sec:structuredjetresults}, rather than uniform jets, are
required to fit the data for both models. Finally, we discuss our
results and further ways these two models might be tested,
\S\ref{sec:discussion}.

\section{The Models}
\label{sec:models}

In this paper we model the observations of the radio nebula from the
Dec. 27, 2004 event. In particular we focus on Epoch II, defined by
the rebrightening of radio flux and proper motion of its centroid and
assume it to be distinct and independent of the emission of Epoch I.
We adopt the natural hypothesis that Epoch II is the result of the
interaction of material ejected during the GF event with an external,
interstellar medium (ISM) \citep{gleg+05,tggg+05,grrte+06}.  Our goal
is to understand and constrain how this interaction might have
transpired. In particular, we attempt to address four key issues {\it
a)} the rapid acceleration of the centroid motion, commencing at $\sim
20$ days, {\it b)} the rapid transition, $\Delta t_{pk} \sim 10$ days,
from rise to decay of the rebrightening peak, {\it c)} the final decay
slope $\propto t^{-1.1}$, and {\it d)} the observed 2:1 aspect ratio
of the elongated emission ellipse, with major axis along the direction
of centroid proper motion.

\begin{figure}[!b] 
\centering
\includegraphics[width=3.5in, angle=-90]{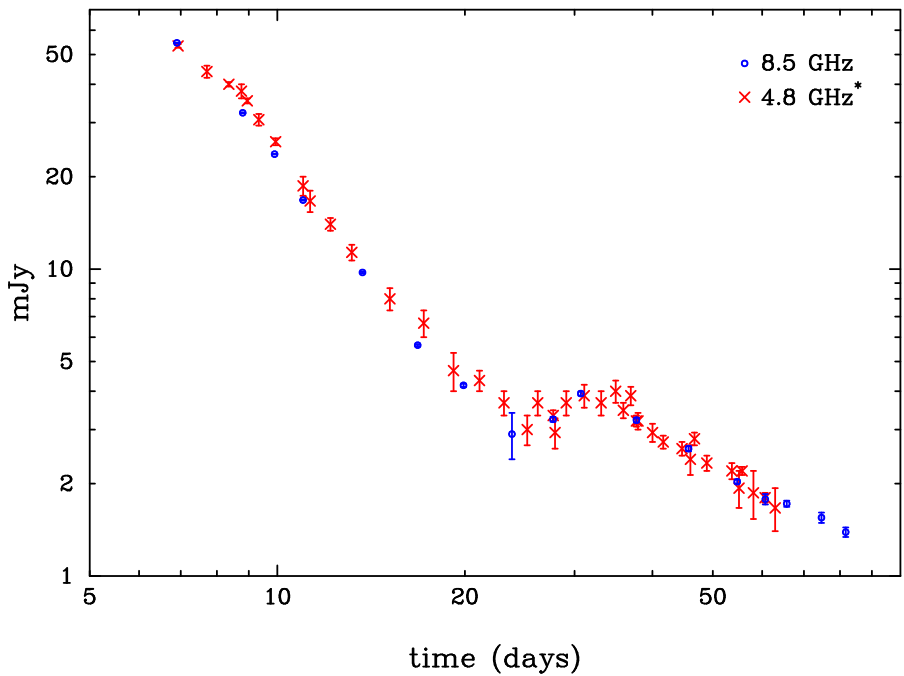} 
\caption{4.8 GHz data \citep{gleg+05} scaled to 8.5 GHz data
\citep{tggg+05} by $F_{4.8}/F_{8.5} = 1.5$, implying $F_\nu \propto
\nu^{-0.6}$.  This scaled 4.8 GHz is denoted with an asterisk ($^*$)
throughout this paper.} \label{fig:p_datcmp}
\end{figure}

To address these issues we begin by assuming two independent flux
components.  The first gives rise to the Epoch I radio flux,
independent of that of Epoch II, and can be described as a stationary
point source at the location of the SGR, empirically parameterized as
a decaying power-law
\begin{equation}
  F_{\mathrm{I}}(t) \equiv 12\ \biggl(\frac{\nu}{8.5 \mathrm{GHz}}\biggr)^{-0.6}  \biggl(\frac{t}{\mathrm{day}}\biggr)^{-2.7}
  \quad\mathrm{Jy} ~,
\label{E:FI}
\end{equation}
where the frequency dependence is motivated by
Figure~\ref{fig:p_datcmp}.  We also assume that this flux component
is present throughout all observations and both epochs, but will be
referred to as the Epoch I flux component.  While invoking
equation~\ref{E:FI} might gloss over the previously mentioned
substructure evident at early times, we argue that this
approximation is adequate and necessary to address its later
interaction with Epoch II.  For example, while there was a
suggestion of proper motion of the radio nebula during Epoch I, it
was not monotonic, but instead the centroid returned to effectively
zero offset; nor was the motion in the same direction as the proper
motion of Epoch II, and thus was not obviously directly related. The
second flux component, designated Epoch II, is entirely derived from
the shock interaction of ejecta emitted during the GF with the ISM.
We wish to test the extent to which this simple two-component model
can describe the data.

The assumption of a two-component model is a departure from that put
forward by \citet{grrte+06}, which proposes that the Epoch I
emission component is due to the collision of the ejecta with the
ISM at some time prior to the first observation at 6.9 days. For the
reasons enumerated above, we interpret Epoch I to be physically
distinct from Epoch II. This assumption is further bolstered by the
observation that the distance the centroid has moved away from the
location of the SGR is correlated with the size of the semi-major
axis along the same direction, suggesting that the SGR continues to
contribute to the flux throughout Epoch II.

Within this two flux component framework, our investigation forks
into two branches.  The first explores collisionally brightened flux
in Epoch II, whereby the material ejected in the GF coasts largely
unhindered until hitting a discontinuous jump or wall in the ISM,
which induces a strong shock and rapid flux brightening.  The second
branch is Doppler brightening, whereby the GF ejecta is thought to
have intercepted the ISM prior to the first observation ($\sim 7$
days) and thus will coast, brightening as it expands $\sim r^2$,
until exhausting its initial kinetic energy to shocked internal
energy, at which time it begins decelerating.  This model is most
like that proposed by \citet{grrte+06} except that they propose the
shock due to the initial collision ($< 7$ days) might account for
Epoch I.  Such a prediction does not naturally flow from our
simulations; although we do model the ejecta colliding with the ISM
at an early time, $r_{wall}/v_0 \approx 6$ days (Table
\ref{tab:data}), and self-consistently calculate any flux deriving
from that collision, we find any such flux to be negligible. Efforts
to increase the ISM wall density, $n_{wall}$, were not
successful in generating a non-negligible early flux component, but
instead would significantly impact the ejecta mass and thus alter
the conditions of the later evolution.  Since the focus of this
paper is on the later evolution of Epoch II, we assume the two flux
components to be independent.

\subsection{Collisional Brightening Model}
\label{sec:collision}

Now we use the empirical emission model for Epoch I
(equation~\ref{E:FI}) to place constraints on the size and shape of
the emission of Epoch II.  At the time of peak brightness of Epoch
II, $t_{pk} \approx 30$ days, the flux centroid was offset from the
SGR position by $d_c \approx 130$ mas.  This is the mean position of
both the flux with zero offset, $d_{I} = 0$, from the SGR, $F_I(t)$,
and the flux from the shocked, ejected material, $F_{II}$
\begin{equation}
  d_c = \frac{d_{II} F_{II}}{F_I + F_{II}} \approx 130 \biggl(\frac{Z_{\mathrm{obs}}}{15 \mathrm{kpc}}\biggr) \ \mathrm{mas}~.
\end{equation}
From equation~\ref{E:FI}, $F_I(t_{pk}) = 1.2$ mJy at 8.5 GHz and
using the data by \citet{tggg+05}, the total flux was $F_I + F_{II}
= 3.9$ mJy. Thus we get an estimate for the observed position of the
ejected shock at peak brightness (when it is beginning to
decelerate)
\begin{equation}
  d_{II} = \frac{F_I + F_{II}}{F_{II}} d_c = 1.4 \times 130\ \mathrm{mas} = 0.52 \biggl(\frac{Z_{\mathrm{obs}}}{15 \mathrm{kpc}}\biggr)\ \mathrm{light-months} ~.
\end{equation}
Thus we estimate the projected velocity of the ejected material across
the sky to be $d_{II}/t_{pk} \approx 0.5 c$.  The projected distance,
$d$, moved by an object over an observer interval, $t_{obs}$, can be
related to the material's true initial velocity, $v_0$, and
inclination angle, $\theta_i$, with respect to image plane ($\theta_i
> 0$ is inclined toward the observer) by
\begin{equation}
  t_{obs} = (1 - (v_0/c) \sin \theta_i) \frac{d}{v_0 \cos \theta_i}
\label{E:obstof}
\end{equation}
and therefore a relationship between the true velocity and inclination
is given by
\begin{equation}
  v_0(\theta_i) = \frac{ c }{  \sin\theta_i + (c/v_\perp) \cos\theta_i  }
\end{equation}
where $v_\perp \equiv d/t_{obs}$.  This equation suggests that if
$v_{\perp} = d_{II}/t_{pk} \approx 0.5 c$, then $v_0 \sim v_\perp$
within $10 \%$ for $0^\circ \lesssim \theta_i \lesssim 60^\circ$, but
increases rapidly to $c$ for inclination angles approaching
$90^\circ$.  Finally, the characteristic true distance from the SGR at
which the ejecta peaked in brightness was
\begin{equation}
  R = \frac{d_{II}}{\cos\theta_i} ~.
\label{E:Rdcosth}
\end{equation}

Another interesting constraint is the swiftness with which the Epoch
II flux rebrightening increased, peaked, and began to decay.  This
entire transition happened within a time span of $\Delta t_{pk} \sim
10$ days $ \sim t_{pk}/3$.  If we assume that the peak of emission
corresponds to a self-similar Sedov-Taylor blastwave reaching its
deceleration radius, then we generally expect $\Delta t_{pk} \sim
t_{pk}$, i.e.~it is difficult to make temporal features shorter than
the blastwave timescale.  In our case we assume the emission derives
from a jet with radial opening angle, $\theta_0$, measured from its
axis.  Thus we may estimate this timescale as simply the timespan
between when the observer first sees the nearest (``front'') edge
and the furthest (``back'') edge of the jet reach a characteristic
radius, $R$.  Using equation~\ref{E:obstof} gives
\begin{equation}
  \Delta t_{pk} = t_b - t_f = \frac{R}{c} (\sin\theta_f - \sin\theta_b) = 2 \frac{d_{II}}{c} \sin\theta_0
\end{equation}
where $\theta_f \equiv \theta_i + \theta_0$ and $\theta_b \equiv
\theta_i - \theta_0$.  Surprisingly, this result is independent of
the derived quantities $R$ and $\theta_i$ and allows one to solve
for the opening angle as a function only of observed quantities
\begin{equation}
  \theta_0 = \arcsin\biggl(\frac{c \Delta t_{pk}}{2 d_{II}}\biggr) \approx 20^\circ \biggl(\frac{0.5 c}{v_\perp}\biggr) \biggl(\frac{15 \mathrm{kpc}}{Z_{\mathrm{obs}}} \biggr)~.
\label{E:theta0}
\end{equation}
If, as has been suggested \citep{tggg+05}, the front of the jet moves
at twice the velocity of the centroid, i.e.~$v_\perp = 2 \times d_c/t_{pk} \sim
0.7 c$, then we expect the opening angle to be $\theta_0 \sim
14^\circ$.

\subsection{Doppler Brightening Model}
\label{sec:doppler}

The rapid peaking of Epoch II can also provide a constraint on the
angle of inclination and ejecta velocity if we assume that this
transition is due only to the ejecta beginning to decelerate as it
moves into, and sweeps up the interstellar medium.  This is distinct
from the model described in the previous section in which the rapid
rise in flux of Epoch II is due to the collision of the ejecta with a
discontinuous rise in external density.  As previously mentioned, due
to the self-similar nature of this expansion, we expect the time to
peak $t_{pk} \propto r_{dec}/v_0 \sim \Delta t_{pk}$, where the
deceleration radius, at which the ejecta has expended its expansion
energy by shocking the external medium, scales like $r_{dec} \propto
[E_0/(n_{ext} v_0^2)]^{1/3}$, where $E_0$ is the ejecta energy and
$n_{ext}$ is the density of the external medium.  If the observed
temporal shortening is assumed to be due entirely to Doppler blue
shift, then the observation $\Delta t_{pk} \sim t_{pk}/3$ implies $(1
- v_0 \sin \theta_i) \sim 1/3$, where $t_{pk} = d/(v_0 \cos
\theta_i)$.  Then using equations~\ref{E:obstof} and \ref{E:Rdcosth}
we have
\begin{equation}
  t_{obs} = (1 - v_0 \sin \theta_i) \biggl(\frac{r_{dec}}{v_0}\biggr) \approx \onethird \biggl(\frac{ r_{dec}}{v_0} \biggr)
\end{equation}
and assuming the proper motion of the head of the jet is twice the value of the centroid
\begin{equation}
  v_\perp = \frac{ r_{dec} \cos \theta_i }{t_{pk}} = \frac{ 2 d_c }{t_{pk}} \approx 0.7 c
\label{E:vperp}
\end{equation}
allows us to solve for the inclination and velocity uniquely
\begin{eqnarray}
\label{E:thetai_v0}
\theta_i &=& 70^\circ \\
v_0      &=& 0.7 c ~. \nonumber
\end{eqnarray}
This allows us to estimate the deceleration radius, $r_{dec} = 2 d_c
/\cos \theta_i \approx 1.7 \times 10^{17} $ cm (eqn.~\ref{E:vperp}) at
an observed distance $Z_{\mathrm{obs}} = 15$ kpc.  Using this
information and the aforementioned estimate of opening angle,
$\theta_0$, we can find the total energy of the ejecta as a function
of external, interstellar density; $\rho_{ext} = n_{ext} m_p c^2$.  To
do so we note that at the deceleration radius the initial kinetic
energy, $E_k$, of the ejecta has been exhausted by shocking the
external medium
\begin{equation}
E_k \approx \frac{\pi \theta_0^2}{3} r^3_{dec} \rho_{ext} c^2 (\gamma - 1) (1 + \Gamma (\gamma-1))
\end{equation}
where the shocked medium is boosted by the Lorentz factor $\gamma
\equiv (1 - v_0^2)^{-1/2}$ and the factor $(1 + \Gamma (\gamma - 1))$
is the specific enthalpy of the shocked external medium
\cite[e.g.][]{afs05} and here we take $\Gamma = 5/3$.  Since we assume
the initial ejecta material is cold, the total energy is related to
the mass as $E_0 = \gamma M_0 c^2$ and the kinetic energy as $E_k =
(\gamma-1) M_0 c^2$, so $E_0 = E_k /(1-1/\gamma)$ and 
\begin{equation}
  E_0 \approx 1 \times 10^{48} \biggl(\frac{\theta_0}{14^\circ}\biggr)^2 \biggl(\frac{n_{ext}}{1~ \mathrm{cm}^{-3}}\biggr)   ~\mathrm{ergs} ~.
\label{E:E_0}
\end{equation}
This is a remarkably simple formula for the total energy of the GF
ejecta and is determined entirely from kinematic arguments. It is
independent of radiative efficiency factors which plague most energy
estimations.  We have numerically simulated a jet with the parameters
of eqn.~\ref{E:thetai_v0}, iterating to find the best fit to the
observations for our adopted radiation parameters (see
\S\ref{sec:syncselfabs}).  Our best fit is J7UHI (Table
\ref{tab:data}).  From the table we find $\theta_0 = 12^\circ$ and
$n_{ext} = 0.003 $ cm$^{-3}$, for which eqn.~\ref{E:E_0} gives $E_0 =
0.23 \times 10^{46}$ ergs, which is about three times lower than the
actual value from the table as taken from the simulation.  Given the
sensitive dependence of $r_{dec}$ on $\theta_i$, this difference in
total energy measurements corrsponds to $\sim 6^\circ$ uncertainty in
$\theta_i$.  This uncertainty is significantly less than the jet
opening angle, $\theta_0$, and thus is within the uncertainty of the
location of maximum emission, due to Doppler effects, on the shock
front.  Therefore we find the simple eqn.~\ref{E:E_0} to be tolerably
accurate and insightful expression for the Doppler brightening model.

\subsection{Structured Jets}
\label{sec:structjets}

A key point of this study is to explore possible jet structures and
their effects on observables.  The physical motivation for this is
clear; nature need not contrive that the material was ejected in the
SGR as a uniform pancake with a distinct edge.  The actual morphology
of the ejecta is an open question and, as it pertains to gamma-ray
burst afterglows, has been studied by several authors
\citep{sg02, rlr02, kg03, jay03, rlsg04}.

We study three basic jet morphologies, all characterized by an opening
angle $\theta_0$ measured from the jet axis.  The ``uniform'' jet is
the simplest, where the ejecta mass density, $\rho$, and initial
velocity, $v_0$, are constants for the region $\theta < \theta_0$, but
are both otherwise zero.  The ``power-law'' jet is constructed by
multiplying the ejecta mass density by the angle factor
\begin{equation}
  \mathcal{A}_P(\theta) \equiv \frac{1}{1 - (\theta/\theta_0)^2}
\label{E:angfac}
\end{equation}
and the velocity is $v(\theta) = v_0 \mathcal{A}^s_P(\theta)$, where
$s$ is typically between $0.5 - 1.0$.  The ``gaussian'' jet multiplies
the mass density by
\begin{equation}
  \mathcal{A}_G(\theta) \equiv \exp \biggl[-\onehalf \biggl(\frac{\theta}{\theta_0}\biggr)^2 \biggr]
\label{E:angfacgauss}
\end{equation}
and the velocity is $v(\theta) = v_0 \mathcal{A}^s_G(\theta)$.
These three jet types are indicated by a ``U'', ``P'', or ``G''
respectively in Table \ref{tab:data} and the power of the velocity
factor, $s$, is unity unless otherwise given in parenthesis.

\section{Numerical Modeling: \Cosmospp }
\label{sec:modeling}

In this paper we report numerical simulations using \Cosmospp, a
recently completed multidimensional,
radiation-chemo-magnetohydrodynamic code for both Newtonian and
(general) relativistic flows. The capabilities and tests of \Cosmospp
are described in \citep{afs05}. This code represents a significant
advance over our previous numerical code \citep[Cosmos;][]{ann03},
with improvements in ultra-relativistic shock-capturing capabilities,
regularization of the symmetry axis, more accurate energy conserving
methods, and adaptive mesh refinement (AMR). The mildly relativistic
flows considered in this paper require accurate relativistic
hydrodynamical capabilites and substantial resolution to resolve the
narrow, length-contracted shock fronts and flow features.  Thus this
present work requires of all of these features and advances.

The simulations are performed using special relativistic hydrodynamics and are
initialized with a cold shell of ejecta mass $M_0$, and total energy
$E_0$ at radius $r_0$ from the SGR moving with an initial
radial speed $v_0$.  Since the ejecta exhibited motion in a particular
direction, we take the shell to be an axisymmetric jet with a
characteristic radial opening angle $\theta_0$ from the jet axis.
This angle, $\theta_0$, is either a distinct edge to a uniform jet, or
might be a characteristic angle over which density and/or velocity
vary as a function of angle from a structured jet axis, as was
discussed in detail in \S\ref{sec:structjets}.  The shell
thickness is constrained to be $\sim 1$ light-hour, corresponding to
the minimum zone size of the refined grid. If the ejecta is linked to
the GF event it was likely ejected on a much shorter timescale,
perhaps seconds, which is consistent with the idea that, since the
ejecta appears to have a specific direction, it must have been ejected
over a fraction of a rotation period, $7.6~s$, of the SGR unless it
was ejected along the rotation axis.  However, there was likely a
dispersion to the ejecta velocity, or a temperature in the gas, so the
shell would likely have expanded significantly over the ensuing days
as it coasted.  Thus our initial shell width is not unreasonable.

The simulations are initiated on a two-dimensional ($r$, $\theta$)
base grid with a typical resolution: $108 \times 24$. The grid
covers the spatial domain $0.01 \lesssim r \lesssim 0.4$
light-years, and $0 \le \theta \le \pi/3,~\pi/4$ or $\pi/6$, with a
reflective boundary at $\theta=0$. The resolution of the base grid
is $\Delta r \approx 1$ light-day.  Four to six levels of refinement
are allowed on this base grid during the evolution. The refinement
and derefinement criteria are described in \citet{afs05} and
use the magnitude, slope, and curvature of the density field to
ensure that the Lorentz-contracted shell of the blast wave is always
maximally resolved, while low density background regions in the flow
are minimally resolved. At maximum refinement, the resolution is
$\Delta r \approx 1$ light-hour or better. The simulations cover a
time span from $t= 5 - 10$ days to $t=75 - 200$ days.

\subsection{Post Processing }
\label{sec:postprocessing}

A key step in assessing the merits of a particular simulation is the
extraction of observable attributes which can be compared with actual
observations. In this case we are interested in simulating
observations at radio wavelengths.  To do this we create synthetic
radio intensity ``images'' for a specified frequency at several
observer times. From these can be obtained an observed flux for that
frequency and time as well as the position and shape of the emitting
region; all of these data were observed for the radio afterglow of
the 27 Dec 04 giant flare and can thus be compared.

\subsubsection{Radiative Transfer }
\label{sec:radtrans}

In \Cosmospp the relativistic hydrodynamic equations are integrated along
parallel time slices, so that for a problem state at a given simulation
(laboratory) time, $t$, all positions, $(x, y, z)$, on the mesh are
simultaneous in time.  In this case the laboratory frame is such that the SGR
at the coordinate origin is stationary and so, to an excellent approximation,
is the terrestrial observer.  If, say, the observer is located at some large
distance $Z_{\mathrm{obs}}$ along the $z$-axis, then he will observe light
from the object that is essentially parallel to the $z$-axis.  Thus the
observer's spatial coordinates $(x_{\mathrm{obs}}, y_{\mathrm{obs}},
z_{\mathrm{obs}})$ perceived simultaneously at time $t_{\mathrm{obs}}$ will
transform from those of the simulation simply as
\begin{equation}
  (t_{\mathrm{obs}},x_{\mathrm{obs}},y_{\mathrm{obs}},z_{\mathrm{obs}})
  = ( t - z/c , x, y, z) - (Z_{\mathrm{obs}}/c, 0, 0, Z_{\mathrm{obs}}) + (r_0/v_0,0,0,0) ~,
\label{eqn:tobs}
\end{equation}
where the last two terms are arbitrary, constant origin offsets.  The
middle term corresponds to look-back time from the Earth and will be
neglected.  The last term accounts for the fact that the simulation is
started (at $t = 0$) after the ejecta, moving at some velocity $v_0$,
has moved a distance $r_0$ from the SGR.

To construct a flux map at a given observer frequency, $\nu$, and time,
$t_{\mathrm{obs}}$, we integrate the radiative transfer equation through
the simulation data, along the z-axis toward the observer
\begin{equation}
  I_\nu(t_{\mathrm{obs}}, x_{\mathrm{obs}}, y_{\mathrm{obs}}, Z_{\mathrm{obs}})
  = \int ( j_\nu - \alpha_\nu I_\nu ) dz_{\mathrm{obs}} ~,
\label{eqn:intensity}
\end{equation}
where $I_\nu$ is the intensity, $j_\nu$ the emissivity, and
$\alpha_\nu$ the opacity of the material at a position
$(t_{\mathrm{obs}}, x_{\mathrm{obs}}, y_{\mathrm{obs}}, z_{\mathrm{obs}})$.
Finally, the observed flux is given by
\begin{equation}
  F_\nu(t_{\mathrm{obs}}, Z_{\mathrm{obs}}) = \int \int I_\nu(t_{\mathrm{obs}}, x_{\mathrm{obs}}, y_{\mathrm{obs}}, Z_{\mathrm{obs}}) \times \frac{ dx_{\mathrm{obs}} dy_{\mathrm{obs}} }{Z^2_{\mathrm{obs}}} ~.
\label{eqn:flux}
\end{equation}
Distance estimates to SGR 1806-20 have been in the range $12 <
Z_{\mathrm{obs}} < 15$ kpc \citep{ce04,fng+05}.  See, however, the lower
estimate of 9.8 kpc obtained from observations of absorption spectra
of this GF event \citep{ccrk+05}.  We take the high value of this
range
\begin{equation}
  Z_{\mathrm{obs}} \approx 15 \mathrm{ kpc}
\label{eqn:distance}
\end{equation}
as a worst case with respect to energy requirements, so that our
reported flux and energy estimates might be relaxed by as much as a
factor $~ (9.8/15)^2 \sim 0.4$, but are not likely to be tightened
(see \S\ref{sec:compare}).

The emissivity and opacity, which are needed in the observer frame,
are transformed from those calculated in the proper (rest) frame of
the fluid, denoted by a prime, as \citep[see eqns. 4.112 \& 4.113
of][]{rl79}
\begin{eqnarray}
  \alpha_\nu &\equiv& \alpha(\nu) = \frac{\alpha'(\nu')}{\delta}
                                 = \frac{\alpha'(\nu/\delta) }{\delta} ~, \\
  j_\nu &\equiv& j(\nu) = \delta^2 j'(\nu')
                       = \delta^2 j'(\nu/\delta) ~. \nonumber
\end{eqnarray}
The proper frequency $\nu' = \nu/\delta$ transforms with the Doppler factor
defined as
\begin{equation}
  \delta \equiv [ \gamma ( 1 - v_z ) ] ^{-1} ~,
\label{E:doppler}
\end{equation}
where $v_z$ is the fluid velocity component along the z-axis (toward
the observer), and $\gamma$ is the Lorentz factor
\begin{equation}
  \gamma \equiv [ 1 - \vec{v} \cdot \vec{v} ]^{-1/2} ~.
\end{equation}
Thus we have a framework to simulate the observed intensity map and
flux for any special relativistic simulation for which the proper
radiation transport quantities $\alpha'(\nu')$ and $j'(\nu')$ can be
determined locally from the state variables of the simulation.  In the
next section we describe this for the case of self-absorbed
synchrotron radiation.

\subsubsection{Self-Absorption Synchrotron Radiation}
\label{sec:syncselfabs}

The observation of non-thermal (power-law) radio emission from the
expanding nebula of this GF suggests synchrotron radiation from the
electrons in the shocked material as the dominant emission mechanism.
The standard theory of self-absorbed synchrotron emission has been
presented in several works \citep[e.g.][and references
therein]{rl79,chev98,lc99}.  Our implementation of this basically
standard model is presented here.  The physical picture of this model
is that hydrodynamic processes such as shocks will energize the plasma
being simulated, thus accelerating its population of electrons into a
power-law distribution as a function of Lorentz factor
\begin{equation}
  n(\gamma) = C \gamma^{-p} ~, 
\label{E:ngamma}
\end{equation}
starting from some minimum Lorentz factor $\gamma_m$.  A key physical parameter
is introduced by assuming that only a proportion, $\epsilon_e$, of the
total proper energy density, $e' = E/\gamma$ from the simulation, is
represented in the electrons.  Thus the fraction of energy in electrons is
\begin{equation}
  \epsilon_e e' = \int_{\gamma_m}^\infty (\gamma - 1) m_e c^2 n(\gamma) d\gamma ~.
\end{equation}
Plugging in equation \ref{E:ngamma}, we can solve for the constant $C$
\begin{equation}
  C = \frac{\epsilon_e e'}{m_e c^2} \biggl[ \frac{\gamma_m^{2-p}}{p-2} - \frac{\gamma_m^{1-p}}{p-1} \biggr]^{-1} ~.
\end{equation}
Another fundamental parameter of the model is the proportion,
$\epsilon_B$, of the total energy density, $e'$, manifest as magnetic
field energy density, $B^2/8\pi$:
\begin{equation}
  B = (8 \pi \epsilon_B e')^{1/2} ~.
\end{equation}

The synchrotron emission results from the population of electrons
moving in, and being accelerated by the $B$-field. The relative
orientation, or pitch angle, of an electron's trajectory with
respect to the $B$-field vector strongly effects the radiation.  We
treat this pitch angle in an average sense by following the
treatment of \citet{wg99}.  They calculate scaling factors such that
the characteristic frequency emitted by a distribution of electrons
(equation~\ref{E:ngamma}) in a field $B$ scales like
\begin{equation}
  \nu'_m \propto \chi_p B ~,
\end{equation}
and the isotropic emission at that characteristic frequency will be
\begin{equation}
  j'(\nu'_m) \propto \phi_p B ~.
\end{equation}
We parameterize their functions of electron index $p$ \citep[see
Fig.~1 of][]{wg99}
\begin{eqnarray}
  \chi_p &\approx& p^{-3/2} + 0.267 ~, \\
  \phi_p &\approx& 0.53 \arctan(p) ~,  \nonumber
\end{eqnarray}
which give values $\chi_p(p = 2.5) = 0.52$ and $\phi_p(p = 2.5) = 0.63$.

So finally, we get for the opacity
\begin{equation}
\alpha'(\nu') = (m_e c^2)^{p-1} \frac{\sqrt{3} q^3}{8 \pi m_e}
                \biggl(\frac{3 q}{2\pi m_e^3 c^5} \biggr)^{p/2}
                C\ (\phi_p B) (\chi_p B)^{p/2}
                \Gamma \biggl(\frac{3 p + 2}{12} \biggr)
                \Gamma \biggl(\frac{3 p + 22}{12} \biggr) {\nu'}^{-(p+4)/2} ~,
\end{equation}
where $\Gamma()$ is Euler's gamma function, $q$ the electron charge,
and we note the factor $(m_e c^2)^{p-1}$ is required to correct the
dimensional inconsistency of the expression in \citet{rl79}.  The
emmissivity is given by
\begin{equation}
  j'(\nu') = \frac{\sqrt{3} q^3 C \phi_p B}{(4\pi) 2\pi m_e c^2 (p+1)} \Gamma\biggl(\frac{p}{4} + \frac{19}{12}\biggr) \Gamma\biggl(\frac{p}{4} - \frac{1}{12}\biggr) \biggl( \frac{2\pi m_e c \nu'}{3 q \chi_p B} \biggr)^{-(p-1)/2} ~.
\label{eqn:jnu}
\end{equation}

The radiative properties of a parcel of shocked material from a
simulation depends on the proper energy and velocity state variables
of the plasma in the simulation, $(e', \vec{v})$, as well as the
emission modeling parameters $(p, \gamma_m, \epsilon_e,
\epsilon_B)$. The minimum electron Lorentz factor, $\gamma_m$, is
typically assumed to be a few.  For the wavelengths primarily
considered here we don't find particular sensitivity to this
parameter, and so deem it sufficient to set it to $\gamma_m \approx 2$.

The electron index, $p$, impacts both the spectral and temporal
decay slopes and therefore it is desireable to have an independent
indication of its value.  An intermediate result of this paper is
that the 4.8 GHz \citep{gleg+05} and 8.5 GHz data \citep{tggg+05}
are consistent with $F_{4.8}/F_{8.5} \approx 1.5$ throughout the
observational period (Figure~\ref{fig:p_datcmp}).  Furthermore, the
single observation by \citet{tggg+05} at 22.5 GHz at $\sim 23$ days
is also consistent with the other points when scaled accordingly.
This implies $F_\nu \propto \nu^{-0.6} \propto \nu^{-(p-1)/2}$
which, in turn implies
\begin{equation}
  p \approx 2.2 ~,
\label{eqn:p22}
\end{equation}
which is the value we use throughout this paper.  While it is clear
that detailed multiwavelength observations yield rich spectral
substructure \citep{ccrk+05} it is also the case that the shorter
wavelengths, where we have the best chance at accurately modeling the
emission, can be represented by a single electron index.

\subsection{Comparison with Data}
\label{sec:compare}

To fit the simulation to the data one must synchronize the
simulation time and observer time by adding $r_0/v_0$ to the
simulation time (equation~\ref{eqn:tobs}) and then determine the
normalization, $\mathcal{E}_N$ (Table \ref{tab:data}), required to
best match the flux magnitudes.  In all of the models presented here
and summarized in Table \ref{tab:data}
(save the deliberate exceptions J5A40SE2, J5A40SE.5, J5A40SEN2) the
simulations were calibrated to match the data as nearly as possible,
so $\mathcal{E}_N$ can be thought of as a fine tuning parameter to
match the data precisely.

Since we find the optical depth of these systems to be negligible, the
magnitude of the flux has the dependence
\begin{equation}
F_\nu \propto j' \propto \mathcal{E}_N
      \biggl( \frac{\epsilon_e}{0.1} \biggr)
      \biggl( \frac{\epsilon_B}{0.1}\biggr)^{(p+1)/4}
      \biggl(\frac{\nu}{\mathrm{8.5 GHz}} \biggr)^{(1-p)/2} {e'}^{(p+5)/4} ~,
\end{equation}
where $e'$ is the proper internal energy of the fluid in the
simulation, $\epsilon_e$ and $\epsilon_B$ parameterize the amount of
this energy in electrons and magnetic field respectively, and
$\mathcal{E}_N$ is the final efficiency or normalization required
to scale the simulated flux to that of the data.  As denoted,
throughout this paper we assume 10\% of the energy is expressed in
each the electrons and magnetic field.

A convenient and robust scaling can be applied to the interpretation
of these simulation results.  As discussed in
\S\ref{sec:modeling}, the initial ejecta energy, $E_0$, is
deposited as a cold gas into a fixed volume (presumably constrained
by the deposition timescale). Thus $E_0 \propto \rho v_0^2$.  Also,
as the ejecta expands and shocks, it will heat according to $e'
\propto \rho v^2$.  For a class of simulations in which a key
hydrodynamic size scale is held fixed, i.e.~the deceleration radius
$r_{dec} \propto (E_0/n/v^2_0)^{1/3}$, we have $n \propto E_0$ for a
given $v_0$ and this enforces similar hydrodynamic evolution, $v
\sim v_0 (r_{dec}/r)^{-3/2}$.  Thus for a given model, the shock
heating scales with total energy; $e' \propto E_0$.  So the
simulated flux scales as
\begin{eqnarray}
\label{E:scale}
  F_\nu &\propto& j' \propto \mathcal{E}_N
         \biggl( \frac{\epsilon_e}{0.1} \biggr)
         \biggl( \frac{\epsilon_B}{0.1}\biggr)^{(p+1)/4}
         \biggl(\frac{\nu}{\mathrm{8.5 GHz}} \biggr)^{(1-p)/2}
         \biggl(\frac{Z_{\mathrm{obs}}}{ 15 \mathrm{kpc}}\biggr)^{-2}
         {E_0}^{(p+5)/4} ~, \\
  n &\propto& E_0 ~,        \nonumber
\end{eqnarray}
where $n$ denotes all external number densities;
$n_{ext},n_{int},n_{wall}$.  For example, models J5A40S and
J5A40SEN2 differ solely in that their total energy, $E_0$, and
external densities $n_{ext}, n_{int}, n_{wall}$ are doubled in
J5A40SEN2 from those in J5A40S.  Equation \ref{E:scale} then
suggests that J5A40SEN2 will have higher flux by a factor $2^{1.8}
\approx 3.5$, using equation~\ref{eqn:p22}.  Indeed, the ratio of
factors $\mathcal{E}_N$ is precisely $1/3.5$ so that both
simulations match the data (see Table \ref{tab:data}).  In fact, we
show below that the resulting flux curves are indistinguishable from
one another. Not only does this validate the hydrodynamic scaling
expected from using a barotropic, $P \propto \rho^\Gamma$, equation
of state, but also allows for convenient reinterpretation of the
data.  For instance, the factor, $\mathcal{E}_N$, can be used to
scale $M_0, E_0, n_{ext}, n_{int},n_{wall}$ by
$\mathcal{E}_N^{4/(p+5)}$, or to scale one of the other terms in
equation~\ref{E:scale}. Similarly, if an independent measurement
were to constrain the external density or efficiency parameters, one
can then calculate that effect on the other physical parameters.
Alternative distances to the SGR (see discussion at
equation~\ref{eqn:distance}) can be accommodated.  The uncertainty
in the emission physics parameters can be addressed by
\begin{equation}
  E_0 \propto n \propto \biggl(\frac{\epsilon_e}{0.1} \biggr)^{-4/(p+5)} \biggl( \frac{\epsilon_B}{0.1}\biggr)^{-(p+1)/(p+5)} ~.
\label{eqn:enscale}
\end{equation}

\subsection{Aspect Ratio}
\label{sec:aspectratio}

One metric that can be used to compare the shape of the observed
radio nebula to that of simulations is the aspect ratio.
Observationally, this value is the ratio of minor to major axes of
an ellipse fitted to the shape of the observed radio nebula.  In
order to make the most realistic comparison with observations
possible, we begin with our simulated flux map $F_\nu$ given by
equation~\ref{eqn:flux} and convolve it with a 2D gaussian filter,
$\propto \exp(-1/2(\varrho/\sigma)^2)$ to blur the image (call it
$F^*_\nu$) where $\varrho \equiv \sqrt{x^2 + y^2}$. Herein we set
$\sigma = 50$ mas which is reasonable in that it blurs out small
features while not significantly impacting the overall structure of
the image.  The aspect ratio of this blurred flux map, $F^*_\nu$, is
then the ratio of the root mean square width to height defined by
\begin{equation}
  \sqrt{\frac{<x^2> - <x>^2}{<y^2> - <y>^2}}
\end{equation}
where
\begin{equation}
  <\xi^i> \equiv \frac {\int\int \xi^i F^*_\nu dx dy }{\int\int F^*_\nu dx dy}
\end{equation}
and $\xi = [x,y]$ and $i = [1,2]$.

\section{Results}
\label{sec:results}

We present our results as a series of fits to observations in
ascending order of initial ejecta velocity, $v_0$.  For the
collisional brightening models, the velocity range we examine begins
at $v_0 = 0.35 c \approx d_c/t_{pk}$ which is roughly the minimum
velocity the ejecta could have had and still account for the
observed centroid offset, $d_c \approx 130$ mas, and continues to
$v_0 = 0.5 c$, for which we find three example solutions.  In the
case of Doppler brightened models, we find higher velocities are
required, specifically we consider models with $v_0 = 0.7c $ and
$0.8c$.

\subsection{Collisional Brightening Results}
\label{sec:collisionalresults}

\begin{figure} 
\centering
\includegraphics[width=4.5in, angle=-90]{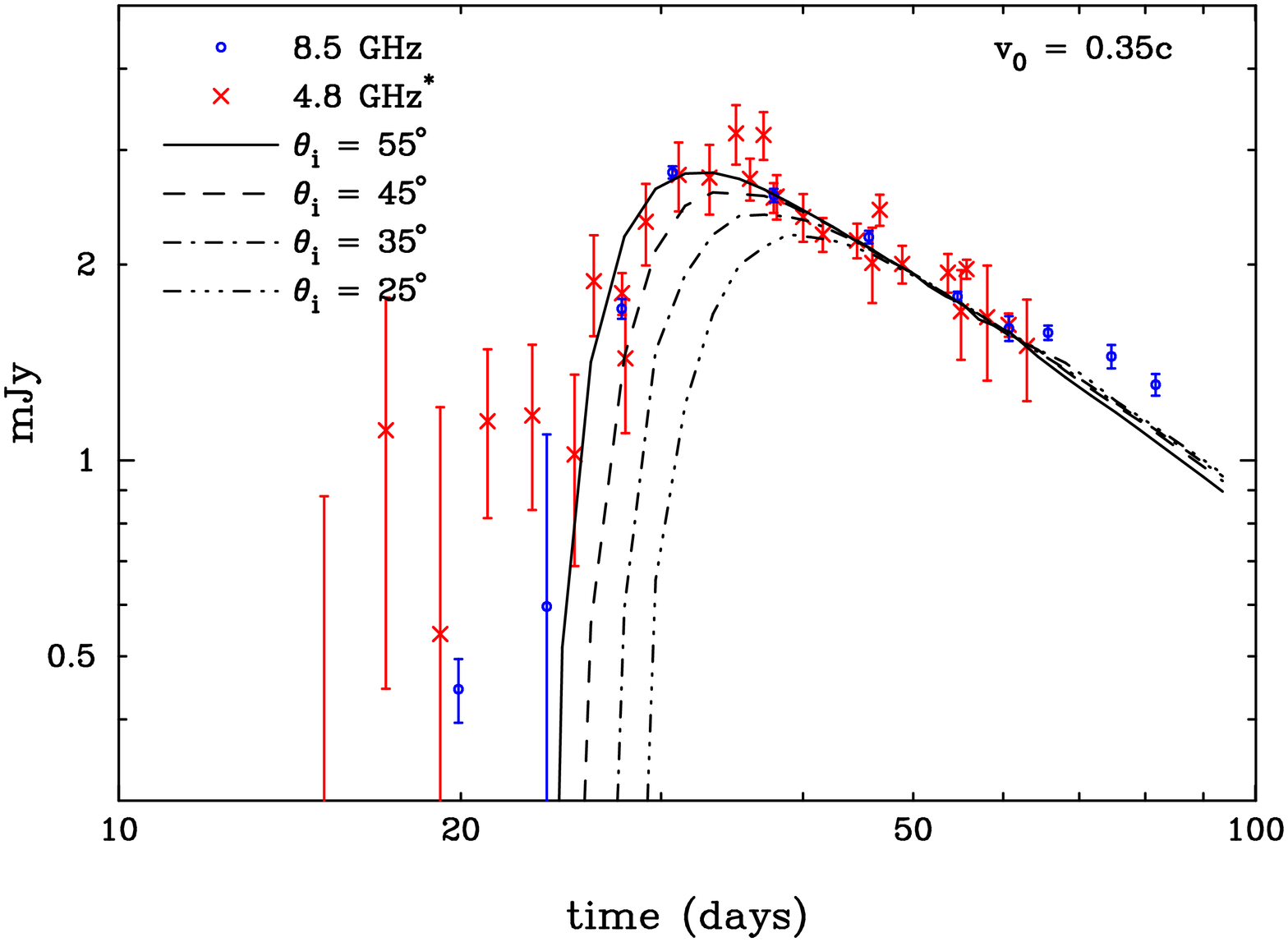} 
\caption{Flux curves of jet collision simulation J35A at four
inclination angles, $\theta_i$, out of the view plane, toward the
observer.  These are compared with 8.5 GHz data \citep{tggg+05} and
scaled 4.8 GHz data \citep{gleg+05}, each with the Epoch I flux
component subtracted off. Inclination angle strongly affects the
timing and brightness of observed flux.  The best fit for this jet
simulation is $\theta_i = 55^\circ$. } \label{fig:v.35_sub}
\end{figure}

At the lowest ejection velocity we consider, $v_0 = 0.35 c$, we
demonstrate the effect on observations of inclination angle,
$\theta_i$, of the jet axis out of the observer view plane.  In
Figure~\ref{fig:v.35_sub} one can see the sensitive dependence of
the Epoch II flux on inclination; the lightcurve is advanced and
doppler brightened as $\theta_i$ is increased.  Note that all curves
fall on the same decay law at late times, as expected due to the
waning importance of doppler effects as the jet decelerates as noted
previously in \citet{jay03}.  By adding the Epoch I component
(equation~\ref{E:FI}) one gets a total light curve and centroid
offset prediction for this simulation, as seen at various
inclinations in Figure~\ref{fig:v.35_tot}.  One can see that either
the flux or centroid off-set curves can be fit reasonably well, but
not both. To do so one must go to higher ejecta velocity.

By increasing the ejecta velocity to $v_0 = 0.377 c$ ($\gamma =
1.08$) one can begin to come close to satisfying both the flux and
centroid offset observations.  In Figure~\ref{fig:v.377_tot} are
shown three separate models, each individually tuned to best match
the data. The tuning is rather crude, but a basic trend is
highlighted. Increasing the density of the wall bounding the ISM,
$n_{wall}$, or the region internal to it, $n_{int}$, causes a more
rapid brightening of the Epoch II flux bump, which can then be
compensated by reducing the inclination, $\theta_i$.  Thus the
morphology of the external medium is degenerate with the inclination
angle.  Note also that model J377I25, with a significant internal
density, $n_{int}$, shows some brightening (and motion of its
centroid) prior to collision with the wall.  This qualitatively
matches the data better and thus argues for a non-negligible
internal medium.

Next we consider $v_0 = 0.5 c$.  While the $v_0 = 0.377 c$ case
nominally satisfied the flux and centroid offset constraints, it was
a bit shy in satisfying the observed centroid offset at 30 days.
Figure~\ref{fig:v.5_tot} shows three individually tuned cases which
each demonstrate good agreement with the data for $v_0 = 0.5c$.  Of
note is the $45^\circ$ deficit in inclination angle between model
J5A355 and the other two. This low viewing angle is compensated for
by reducing the wall radius, $r_{wall}$ and increasing its density
$n_{wall}$ so the jet is seen to collide and brighten at the same
time as the other models.  The other models, J5A40S and J5A40G,
contrast our two basic structural morphologies of the jet, with
angular ejecta mass density variation described by
equations~\ref{E:angfac} and \ref{E:angfacgauss} respectively.  It
is clear that each model can be tuned to fit the observations, thus
making differentiation between the models impossible with present
data. Figure~\ref{fig:v.5_sub} shows several different variation of
model J5A40S, as well as uniform jet model J5A40U.  The striking
agreement of J5A40SEN2 and J5A40S speaks to the robust hydrodynamic
scaling exhibited in these simulations, as discussed in
\S\ref{sec:compare}.  Varying the total energy $E_0$ by a factor 0.5
(J5A40SE.5) or 2 (J5A40SE2) gives an idea of how sensitive these
solutions are to this parameter.  The former(latter) case
reduces(increases) the deceleration radius and thus
increases(decreases) relative flux at early time.  Finally, J5A40U
demonstrates the stark difference in the uniform jet evolution to
the other, structured jets.  This will be discussed further in
\S\ref{sec:structuredjetresults}.

In order to compare the observational features of different models,
it is useful to construct a synthetic, observer-eye-view flux map of
the emission, as shown in Figure~\ref{fig:plotA}.  These images show
the position and shape of the Epoch II emission with respect to the
SGR. Note that all of the non-uniform models exhibit a
characteristic double brightspot at later time - the leading (top)
spot corresponding to the external shock, while the trailing spot
corresponds to the collision of the flow-focused ejecta
(\S\ref{sec:structjets}). Figure~\ref{fig:plotS} shows the same data
as Figure~\ref{fig:plotA}, but with the image convolved with a 50
mas standard deviation gaussian point spread function.  This
blurring is meant to simulate the limitation in observable
resolution.  At this resolution the double brightspots are blurred
together, but a marginal distinction can be made between the Epoch I
and II emission regions.

Taking the images of Figure~\ref{fig:plotS} as being the most
realistic, one can compare the predicted aspect ratio to that of
observations as discussed in \S\ref{sec:aspectratio} and shown
in Figure~\ref{fig:aspectv.5}.  The simulations exhibit a signature
decrease-inflection-increase behavior which is the result of the
emergence of the Epoch II flux component offset from the SGR
(prompting a decrease in aspect ratio) until it surpasses the total
flux from Epoch I (the inflection) and continues to gain dominance
over it (the increase).  The simulations show qualitative agreement
with the data.  Indeed, the hint of a decrease-inflection-increase
behavior in the data, centered at around 30 days or so, is
indicative of the veracity of the basic model of the emission being
comprised of two independent epochs. The rapid decrease in aspect
ratio at $\sim$ 30 days corresponds well with the rapid brightening
of the flux curve and rapid motion of the centroid at this time
(e.g. Figure~\ref{fig:v.5_tot}), indicating that the emergence of
the Epoch II flux component was swift, $\lesssim 10$ days.

\subsection{Structured Jet Results}
\label{sec:structuredjetresults}

We find that the uniform jet model fails to produce the $t^{-1.1}$
decay slope observed for Epoch II, but instead decays more steeply.
This is unexpected when one considers that the flux scaling for a jet
with constant opening angle, $\theta_0$, and expanding with a
Sedov-Taylor blastwave profile should go like
\begin{equation}
  F \propto (r \theta_0)^2 r j' \propto r^3 {e'}^{1.8} 
    \propto r^{-2.4} \propto t^{-0.96} ~,
\label{eqn:funiformold}
\end{equation}
where $(r \theta_0)^2 r$ is the volume of radiating electrons with
emissivity, $j' \propto {e'}^{(p+5)/4}$ (equations~\ref{eqn:jnu} \&
\ref{eqn:p22}).  For a strong shock the proper energy goes like $e'
\propto v^2$ and for a Sedov-Taylor solution, $v \propto r^{-3/2}$,
while finally, $t \propto r/v \propto r^{5/2}$
(equation~\ref{E:obstof}). Indeed we find these scalings to be good
approximations of the simulations - except that $\theta_0$ is assumed
to be constant.  In reality, a rarefaction wave will propagate inward
(i.e.~poleward) from edge of the jet at the sound speed, $c_s$, thus
reducing the effective radiating area.  This rarefaction wave moves
like $c_s \propto \sqrt{P/\rho} \propto \sqrt{e'} \propto v$, where $P
= (\Gamma-1)e'$ and $\rho \propto \mathrm{constant}$ for strong
shocks.  So the jet's radiating surface is roughly constant in time as
it expands, and so the flux goes like
\begin{equation}
  F\propto r j' \propto r^{-4.4} \propto t^{-1.76}
\label{eqn:funiformnew}
\end{equation}
which is consistent with the minimum (i.e.~best fit) asymptotic slopes
of uniform jets that we've simulated.  Therefore, on theoretical and
numerical grounds, the uniform jet does not satisfy the late time flux
curve.

By contrast, both the structured and gaussian jets can be made to
satisfy the $t^{-1.1}$ decay slope. The reason is a phenomenon we call
``flow focusing'', which we find to be quite prevalent in models for
which the velocity decreases with angle from the jet axis.  The basic
mechanism is as follows: If the jet is given some initial velocity
variation $v(\theta)$ decreasing with angle, the ejecta front will
distort as the material in the jet core begins to lead that in the
wings. Thus, as the ejecta drives a shock into the external medium, a
reverse shock is also driven back into the ejecta which, due to its
distortion, will have obliquity and thus impart a poleward velocity
into the ejecta. A secondary collision then results as the shocked
ejecta collides onto the jet axis.  This second shock produces a
region of high pressure behind the on-axis portion of the external,
forward shock which, in turn, drives it outward.  Thus the on-axis
forward shock velocity will be sustained for longer than expected in a
typical Sedov-Taylor blastwave, and thus can create slower observed
flux decay rates.  An example of the flow-focusing in model J5A355 is
shown in Figures~\ref{fig:sgr7i8nb122D} and
\ref{fig:sgr7i8nb12_lo}. One can see a characteristic bowling pin
shape to the forward shock at late time due to the added pressure of
the secondary shock of the ejecta colliding on axis. This simulation
can be contrasted with that of uniform jet J5A40U in
Figure~\ref{fig:figsgr7i8nb17ce22D}, where the lack of obliquity in
the reverse shock does not prompt an on-axis secondary shock and so
the evolution is much more akin to that of a Sedov-Taylor
blastwave. The effect of flow-focusing on the lightcurve can be seen
in Figure~\ref{fig:v.5_sub} where the the uniform jet shows much more
rapid flux decay than the structured jets.

\subsection{Doppler Brightening Results}
\label{sec:dopplerresults}

The last set of models we study here do not invoke a discontinuous
increase in ISM density with a wall to prompt the observed rapid
brightening of Epoch II.  Instead, these models assume a
sufficiently high Doppler factor (equation~\ref{E:doppler}) to
brighten and temporally compress the observed flux curve.  In
general, these models require higher velocities and inclination
angles in order to achieve sufficient Doppler factors, as discussed
in \S\ref{sec:models}.  In particular, we take the estimates from
equations~\ref{E:theta0} and \ref{E:thetai_v0} and seek a uniform
jet model with $v_0 = 0.7c$, $\theta_i = 70^\circ$ and $\theta_0 =
12^\circ$.  The results are shown in Figures~\ref{fig:v.7u_sub} and
\ref{fig:v.7u_tot}.  In general, these models have a much larger
deceleration radius $\propto (E_0/n_{ext})^{1/3}$ (Table
\ref{tab:data}) than the collisional models.  The qualitative
agreement within errors is acceptable. However, one might ask if it is
possible to improve the fit by creating a faster rise to peak and a
slower decay therefrom. Unfortunately, these two goals are at odds
with each other. Obtaining a steeper rise to peak requires a larger
$v_0$ or $\theta_i$ in order to increase the Doppler factor and
expedite the observed evolution. Such an adjustment would, in turn,
also increase the decay slope, which is already steeper than the
observation.

Figure \ref{fig:v.7u_view} shows simulated flux maps similar to
those of Figure~\ref{fig:plotA}.  Because the external density,
$n_{ext}$ is much lower than that of the previously discussed models
($< 1$\%), the forward shock travels further and becomes larger
before decelerating. Figures~\ref{fig:v.7u_view} \& \ref{fig:v.7u_sub}
also demonstrate the level of numerical convergence of this model.
Models J7UHI and J7ULO begin identically (requiring high resolution
initially to resolve the problem features), but the latter
progressively sheds its five layers of mesh refinement (due to an
artificially high derefinement threshold in the proper mass
density), having entirely deresolved the forward shock at the
mid-time of the simulation, until it is entirely on the base mesh at
the end of the simulation.  The remarkably good agreement of the
data from these two simulations is indicative of the stability of
the results.  Figure~\ref{fig:v.7u_t.6} compares the proper internal
energy $e'$, representative of shock heating, for both problems at
the end of the run: 0.6 year.  The problems exhibit excellent
qualitative agreements.

In Fig.~\ref{fig:v.7smear}, the images of Fig.~\ref{fig:v.7u_view} for
model J7UHI are convolved with a 50 mas gaussian as was done in
Figure~\ref{fig:plotS}.  Once again, at this level of blurring, each
epoch's flux component should be resolved. The aspect ratio of this
model is shown in Figure~\ref{fig:aspectv.7}. The more gradual
increase in flux from Epoch II in these uniform ISM models
corresponds to a smoother minimum in aspect ratio (at $\sim 20$
days) than the sharp inflection seen in Figure~\ref{fig:aspectv.5}.

Finally, we explore the possibility of a non-uniform jet plowing
into the uniform, low-density ISM to see what effect flow-focusing
observed in the earlier models might have in this case.  The
lightcurve for such a model, J8S (Table \ref{tab:data}), is shown in
Figure~\ref{fig:v.7last_tot} for two inclination angles, $60^\circ$
and $70^\circ$, and the aspect ratio for these is shown in
\ref{fig:aspectv.7}.  Because of the relatively long distances
traversed by the ejecta, a rather mild velocity variation was
employed, $s = 0.25$ (see equation~\ref{E:angfac}).  It is evident
that J8S does not show improvement over J7UHI in the sharpness of
the Epoch II peak, even though a higher initial velocity, $v_0 =
0.8c$, was used to attempt to increase the Doppler factor.  However,
it does exhibit a more gradual decay slope, $\propto t^{-1.1}$,
compared to $\propto t^{-1.7}$ for J7UHI, which is consistent with
the prediction of equation~\ref{eqn:funiformnew} and is more consistent
with the observed slope.

\section{Discussion}
\label{sec:discussion}

In this paper we have numerically modeled the radio nebula of the GF
of SGR 1806-20 under the assumption that it was comprised of two
independent flux components: Epoch I assumed to be centered on the
SGR and to uniformly decay throughout the observational period; and
Epoch II assumed to be due to the interaction with the ISM of the
mass ejected from the SGR during the GF.  In general we find that
this two component model can satisfy all of the observations
considered here: the lightcurve, the centroid position, and the
nebula aspect ratio.

\begin{figure}
\includegraphics[width=4.5in, angle=-90]{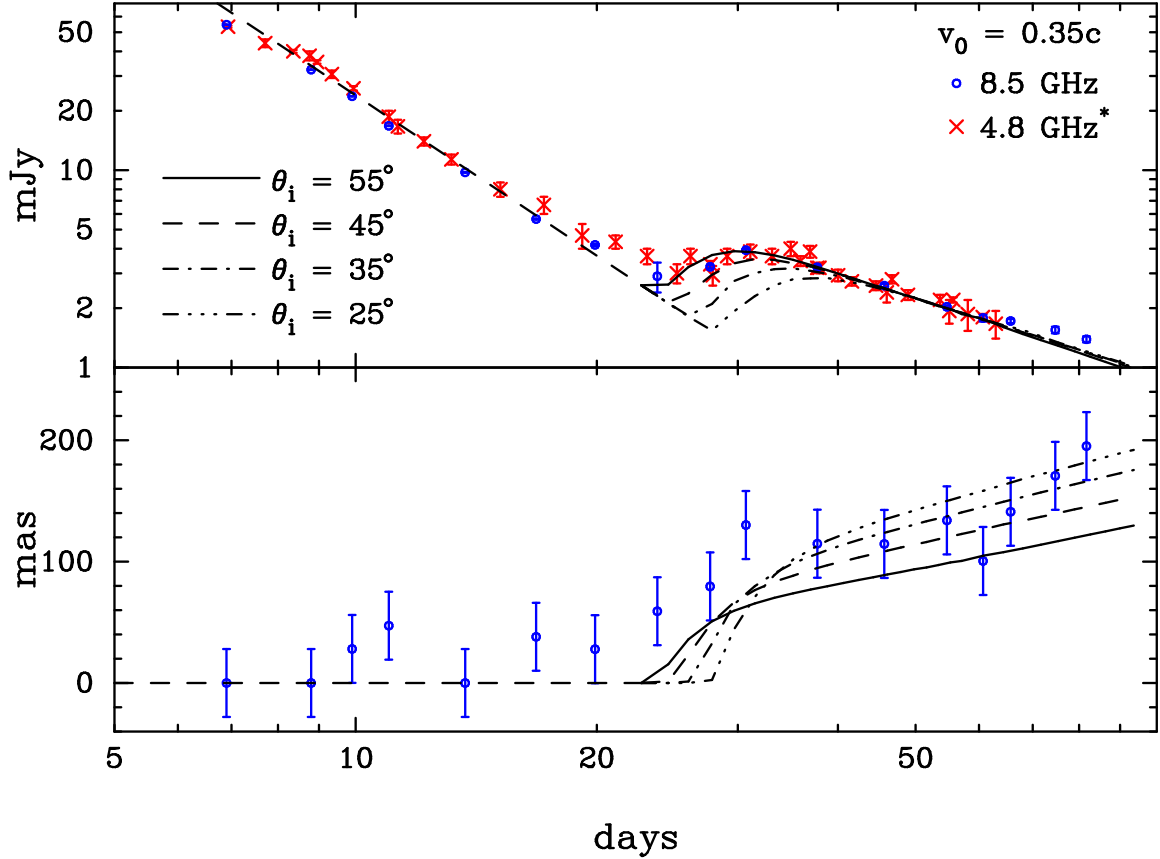} 
\caption{Flux curves (top) and centroid position offsets in
milliarcseconds (bottom) for simulation J35A seen at four
inclination angles, $\theta_i$.  Compare with
Fig.~\ref{fig:v.35_sub} with the same data, but without the Epoch I
flux component. Jet inclination $\theta_i = 55^\circ$ most
accurately matches the lightcurve.  The flux centroid has zero
offset from the SGR until the shock from the ejecta collision with
the external medium brightens. Thus we see a rapid acceleration of
the centroid position between 20 and 30 days.  For inclination
$\theta_i = 55^\circ$, initial ejecta velocity $v_0 = 0.35 c$ is
seen not to be sufficient to produce the observed centroid offset,
but the qualitative motion is evident.} \label{fig:v.35_tot}
\end{figure}

\begin{figure}[htb]
\includegraphics[width=4.5in, angle=-90]{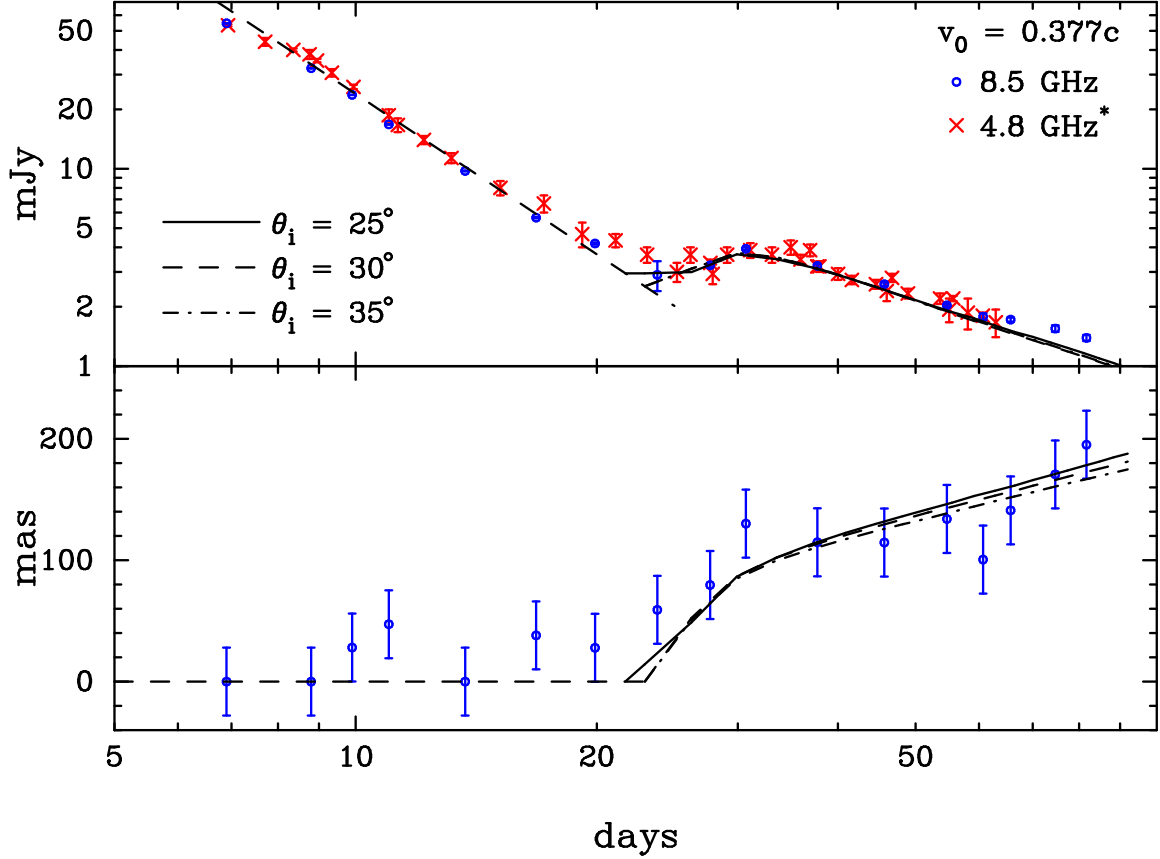} 
\caption{ Flux and centroid offset curves for three simulations,
J377I25, J377I30, J377I35.  All three match the data quite well, but
are individually calibrated to match at individual angles of
inclination, $\theta_i = 25^\circ, 30^\circ, 35^\circ$ respectively.
Model J377I35 is the same, except for $v_0$, as J35A in
Fig.~\ref{fig:v.35_tot}.  Model J377I30 uses higher densities in the
wall, $n_{wall}$ to produce the observed rapid rebrightening at a
shallower inclination.  Model J377I25 adds internal density,
$n_{int}$, thus causing an earlier rise to the Epoch II flux, which
better represents both the flux and centroid position data. }
\label{fig:v.377_tot}
\end{figure}

\begin{figure}
\includegraphics[width=4.5in, angle=-90]{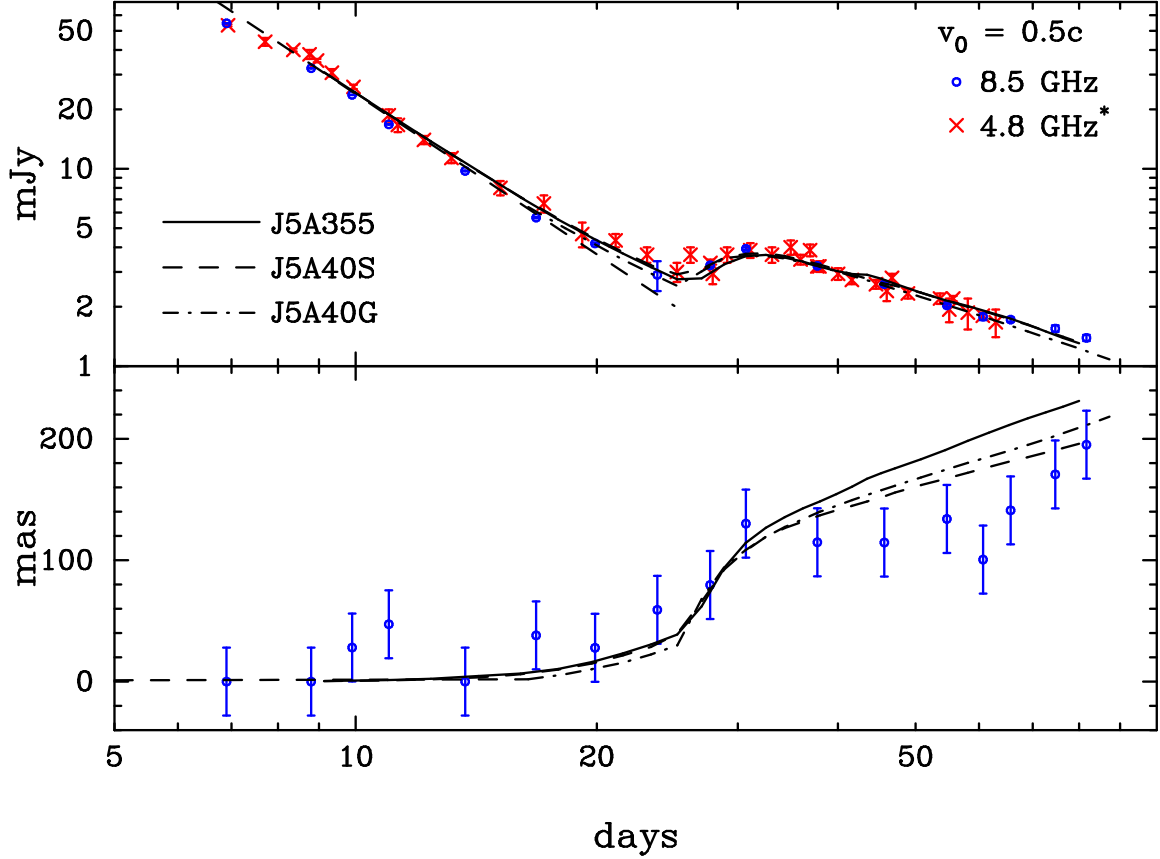} 
\caption{ Flux and centroid offset curves for three simulations at
$v_0 = 0.5 c$: J5A355, J5A40S, J5A40G. These simulations are viewed
at inclination angles, $\theta_i = -5^\circ$, $40^\circ$ and
$40^\circ$ respectively.  Each of these models show excellent
agreement with the data, despite wide variations in the position of
their external wall of material and jet structures.  These models
are typified by ejecta velocity, $0.5c$, which allows consistency
with centroid offset data (bottom plot), a collision with an
external wall initiating the rapid flux rebrightening, and a
structured jet which enables flow focusing and thus reproduction of
the late-time $t^{-1.1}$ flux decay law. } \label{fig:v.5_tot}
\end{figure}

\begin{figure}
\includegraphics[width=4.5in, angle=-90]{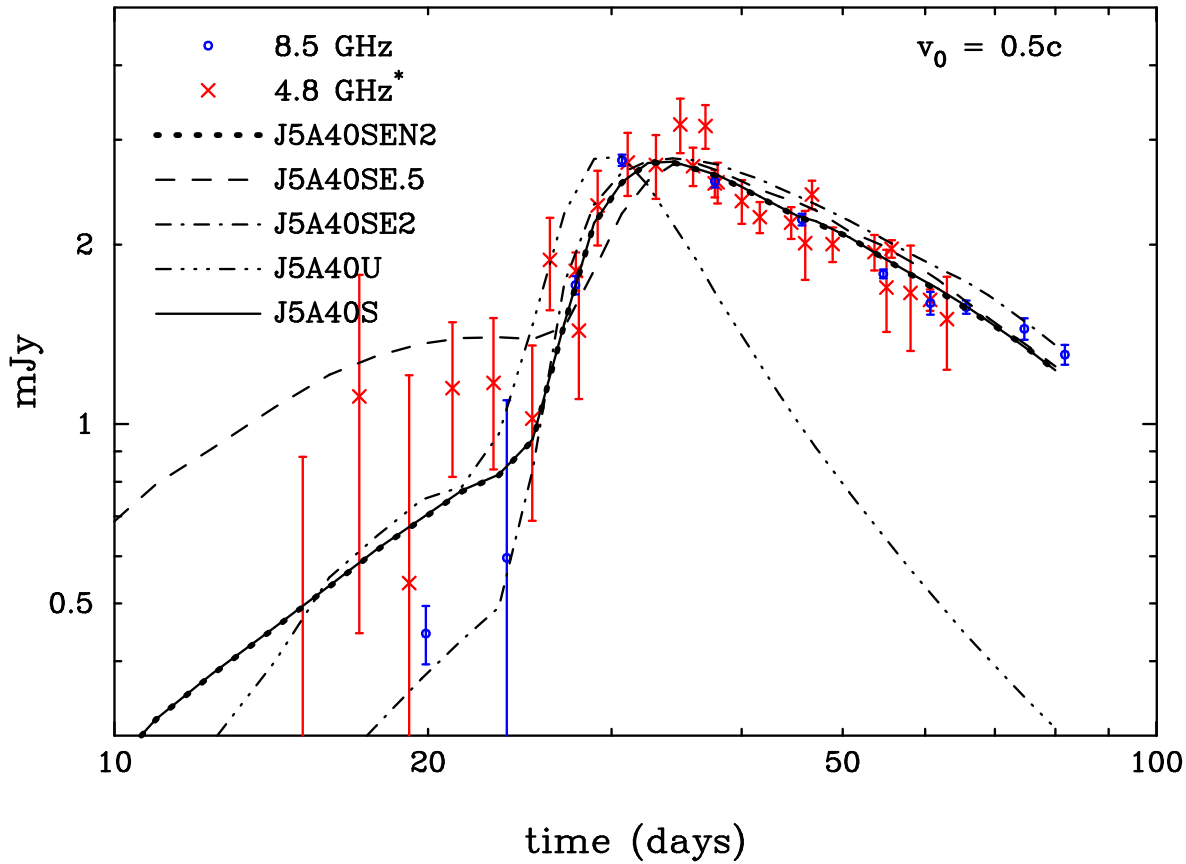} 
\caption{ Flux curves for a range of models for $v_0 = 0.5c$. Model
J5A40S shows good agreement with the data (see
Fig.~\ref{fig:v.5_tot}).  Model J5A40U is a uniform jet calibrated
to the same peak flux.  The rapid flux decay of this model
demonstrates the important role of flow focusing of structured jets
in producing the observed $t^{-1.1}$ flux decay law at late time.
Models J5A40SE2 and J5A40SE.5 compare the effect of doubling and
halving, respectively, the total ejecta energy, $E_0$.  Finally, for
J5A40SEN2, total energy, $E_0$, and all external densities,
$n_{ext}$, $n_{int}$, $n_{wall}$ have been doubled.  The fact that
the flux curve for J5A40SEN2 can be scaled to match that of J5A40S
demonstrates a robust hydrodynamic scaling. } \label{fig:v.5_sub}
\end{figure}

\begin{figure}[htb]
\centering
\includegraphics[width=6in]{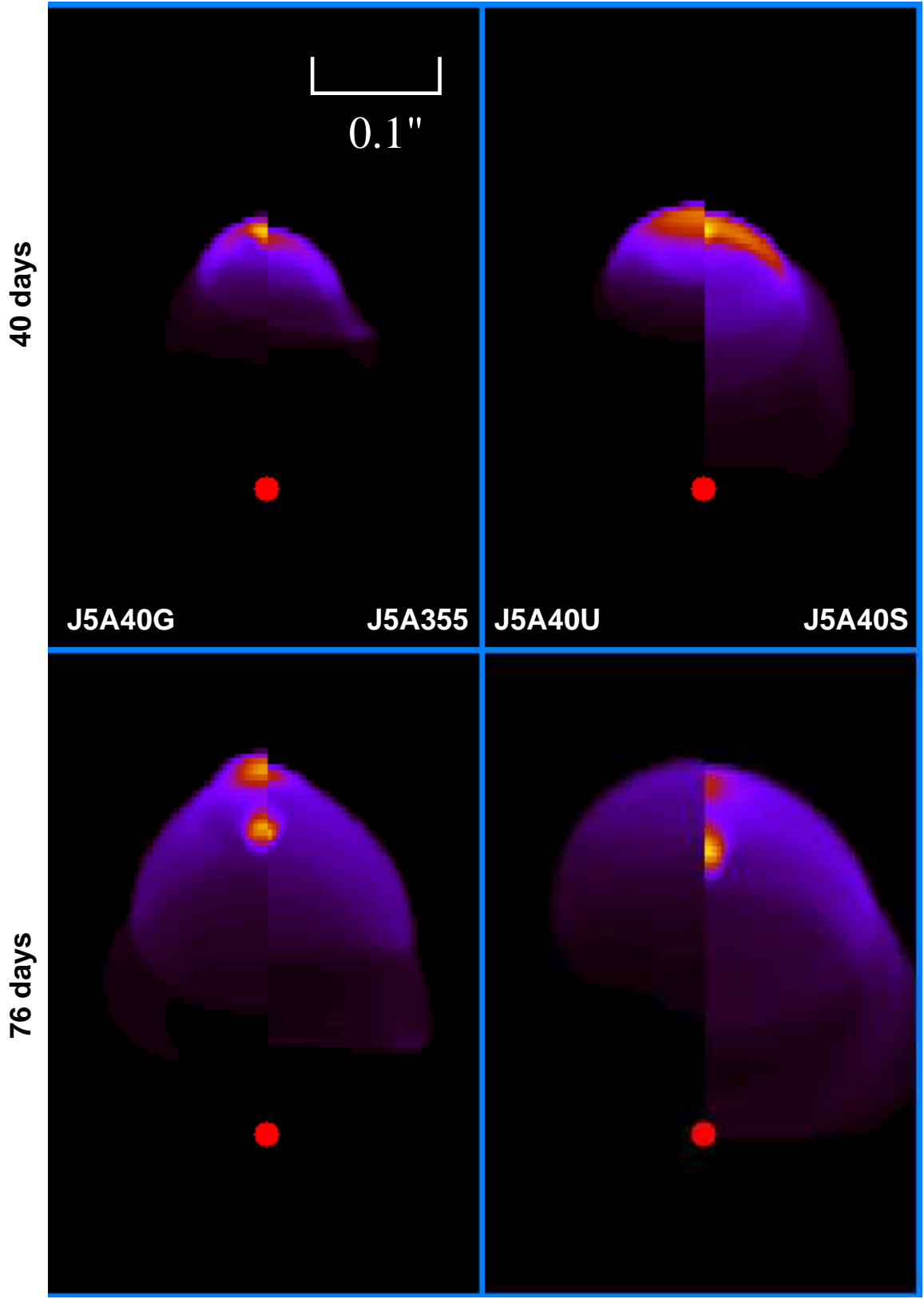} 
\caption{ Simulated flux maps of four models from
Figs.~\ref{fig:v.5_tot} \& \ref{fig:v.5_sub} at two times.  For the
three structured jets one can see the two bright spots characteristic
of flow focusing, while the uniform jet, J4A40U, can be seen to fade
rapidly.  Each of the four frames is $0.08 \times 0.12$ ly$^2$ in size
and both halves of each frame are plotted on the same color scale. }
\label{fig:plotA}
\end{figure}

\begin{figure}
\centering
\includegraphics[width=6in]{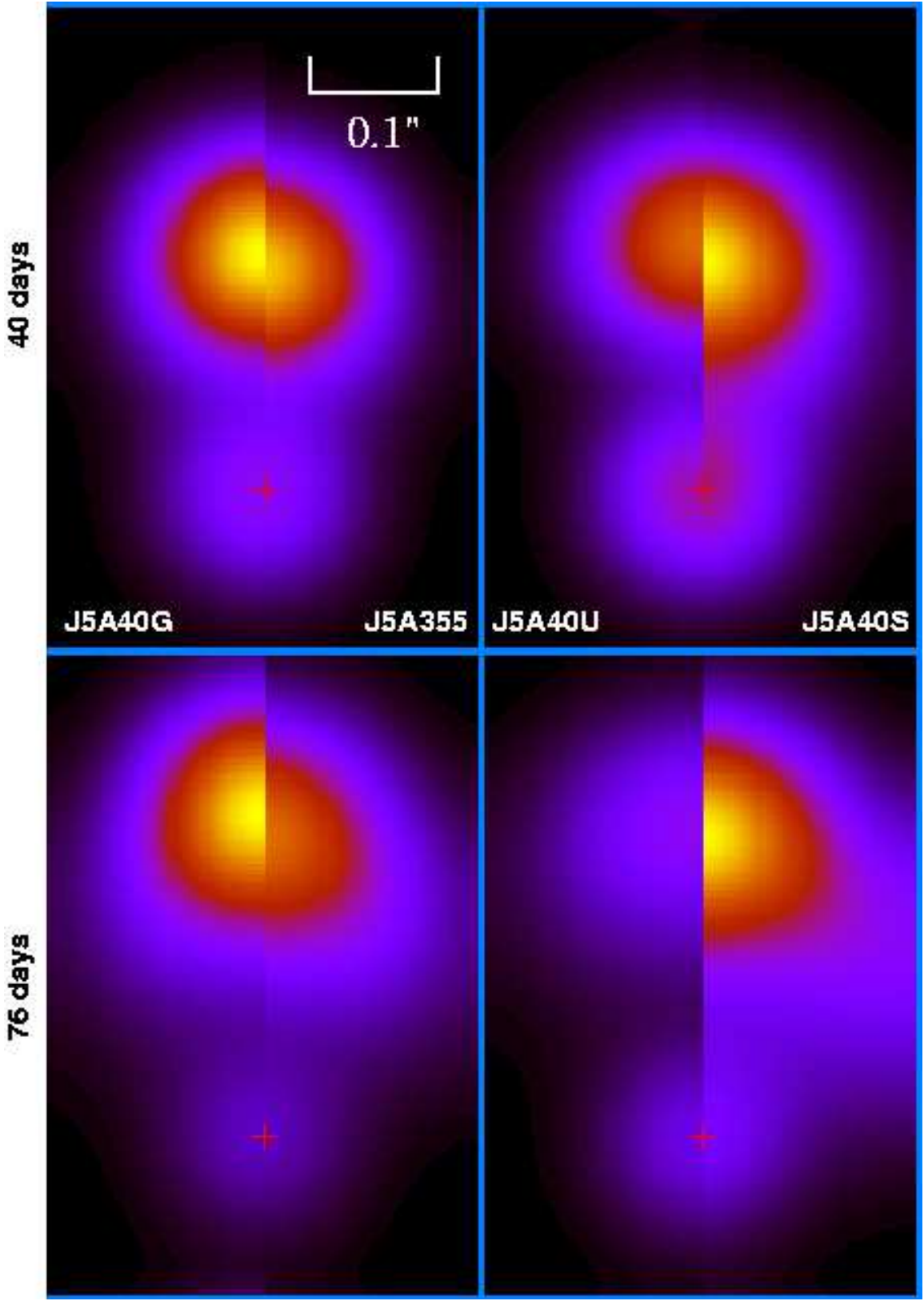} 
\caption{ Same data as the previous figure \ref{fig:plotA}, but with
the Epoch I (eqn.~\ref{E:FI}) flux component added at the position of
the SGR (denoted by the red ``+'') and the image is then convolved
with 50 mas gaussian to simulate resolution limitations.  Thus the
total flux from one of these flux maps corresponds to the total flux
in Fig. \ref{fig:v.5_tot}.  One can see the rapidly fading Epoch I SGR
flux component and that 50 mas resolution begins to differentiate the
two flux components. } \label{fig:plotS}
\end{figure}

\begin{figure}
\includegraphics[width=4.5in, angle=-90]{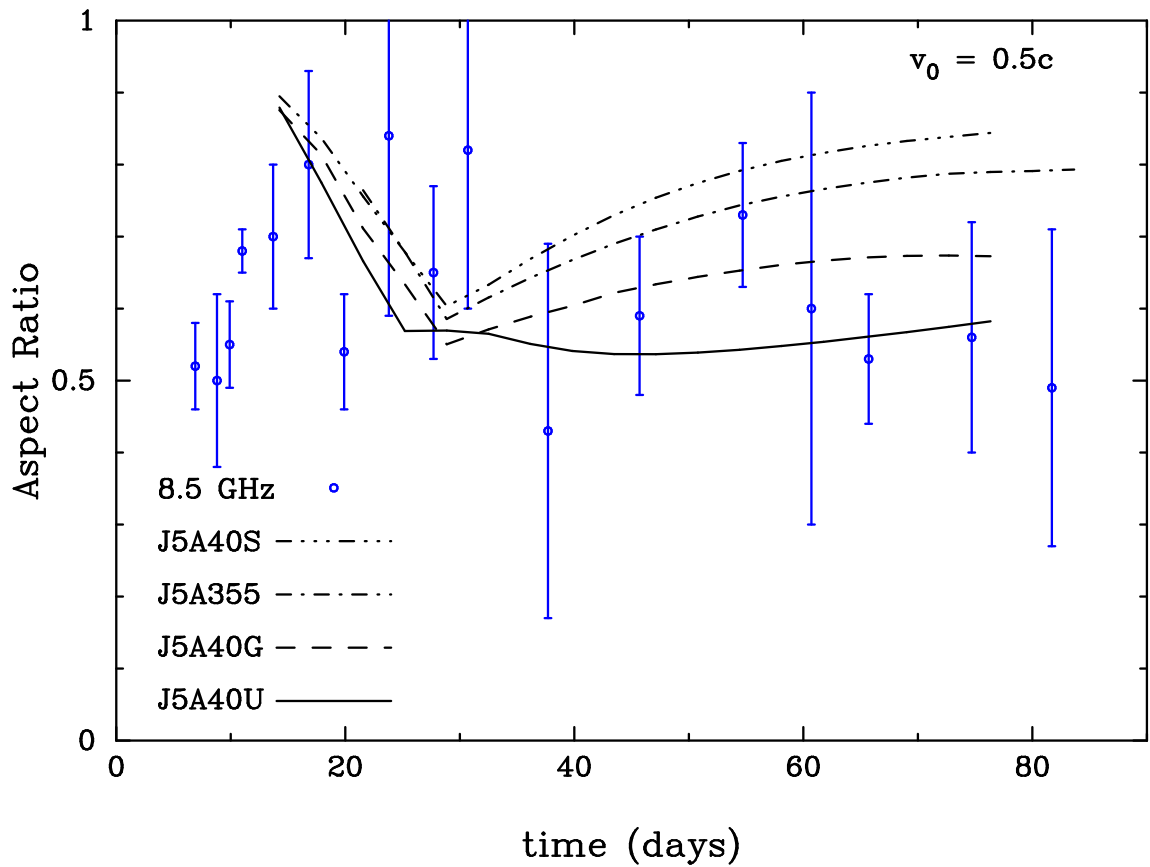} 
\caption{ Aspect ratios are plotted for the simulated flux maps of
the four models shown in Fig.~\ref{fig:plotS}. The simulated aspect
ratio is defined in \S\ref{sec:aspectratio}. This is plotted against
the axial ratio of the ellipse fit by \citet{tggg+05} to the
observed emitting region. All simulations exhibit a decrease in
aspect ratio as the Epoch II flux component increases, and an
inflection and subsequent increase when the Epoch II flux exceeds
that of Epoch I. Model J5A40U shows the most constant aspect ratio
after 30 days because, as seen in Fig.~\ref{fig:v.5_sub}, Epoch II
decays rapidly in this model and thus stays comparable in magnitude
to Epoch I.  It is intriguing that the data tends to hint at the
same decrease-inflection-increase evolution in aspect ratio as the
simulations, indicative of two independent epochs. }
\label{fig:aspectv.5}
\end{figure}

\begin{figure}
\plotone{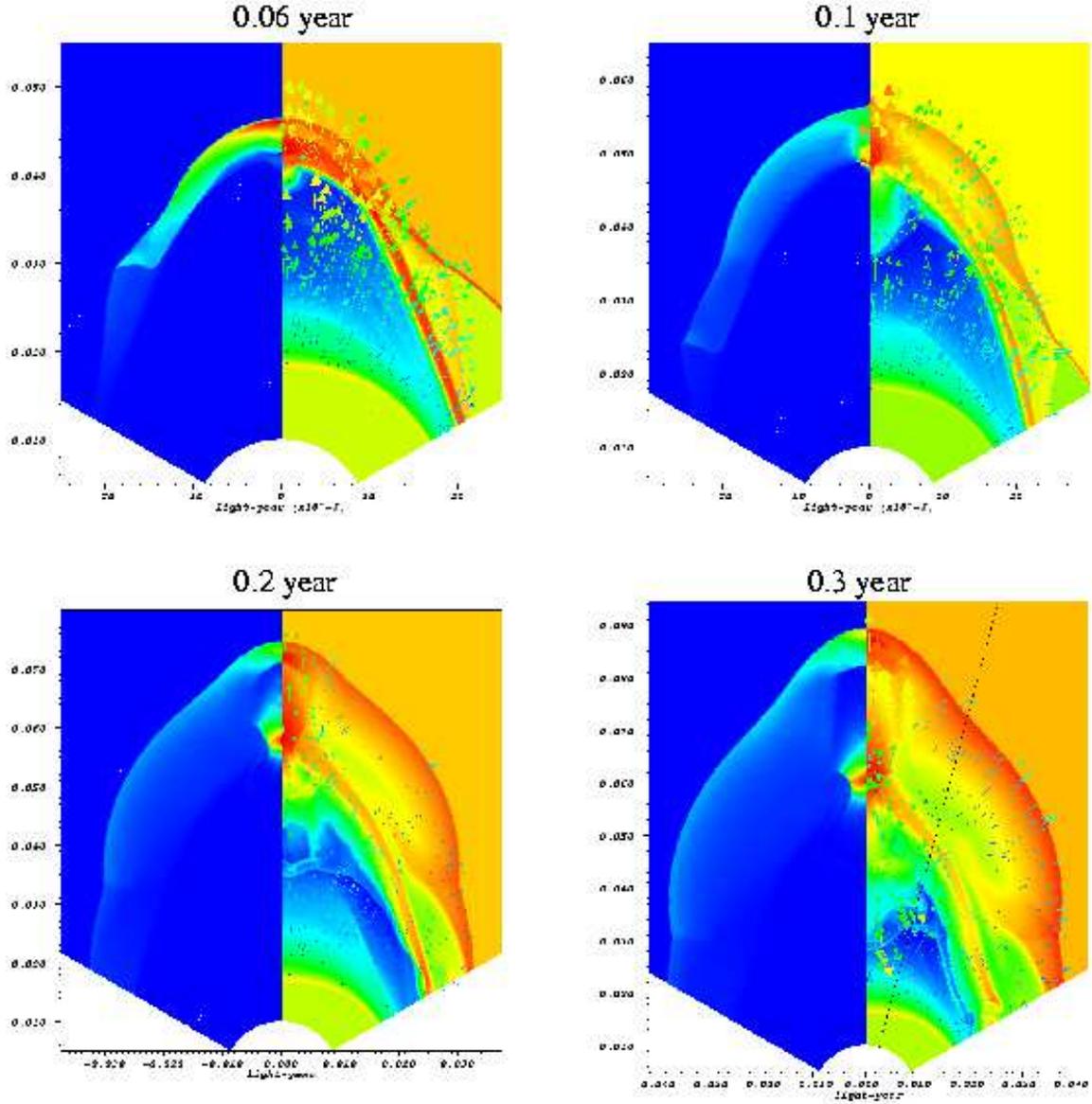} 
\caption{ Snapshots from simulation J5A355 at four times.  On the left
of each panel, the proper internal energy $e'$ is shown.  On the right
half $\log(\rho)$ highlights structure and the coordinate velocity
field $\vec{v}$ is superposed. The ejecta collides with external
medium and its flow is focused toward the jet axis.  This, in turn,
creates a secondary hot-spot on axis that powers the forward shock,
creating the characteristic bowling pin shape of the forward shock at
late times.  At time 0.3 year the 15$^\circ$ lineout (see
Fig.~\ref{fig:sgr7i8nb12_lo}) is shown for reference.  }
\label{fig:sgr7i8nb122D}
\end{figure}

\begin{figure}
\includegraphics[width=4.5in, angle=-90]{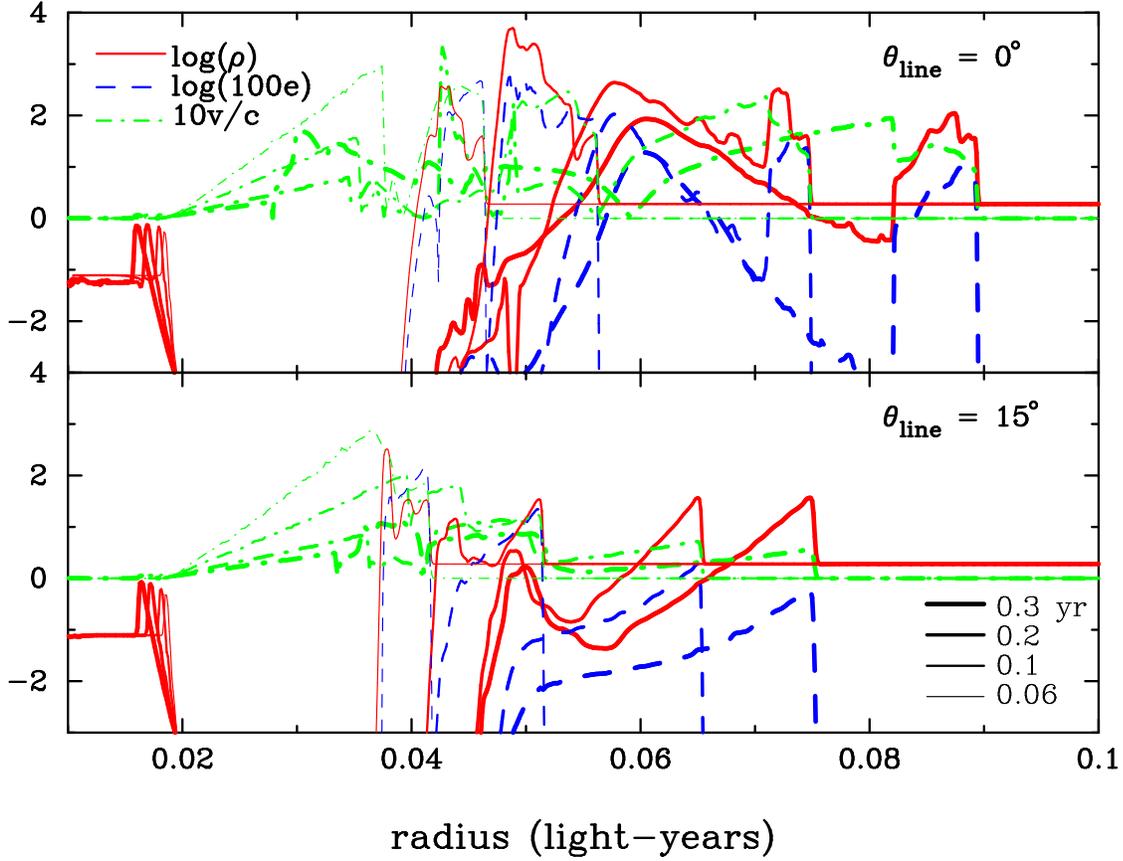} 
\caption{ Radial lineouts through simulation J5A355 along two angles
of inclination, $\theta_{\mathrm{line}}$, from the polar axis.  Proper
density, energy and velocity are represented for the same four times
as shown in Fig.~\ref{fig:sgr7i8nb122D}.  One can see the flow
focusing of original ejecta material onto the jet axis,
$\theta_{\mathrm{line}} = 0^\circ$, at roughly a radius of 0.05 - 0.06
lt-yr, which then drives the center of the jet forward.  The forward
shock can be seen to progress outward more rapidly for
$\theta_{\mathrm{line}} = 0^\circ$ along a time sequence of radii
(0.045, 0.055, 0.075, 0.09) lt-yr, compared with (0.041, 0.051,
0.065, 0.075) lt-yr for $\theta_{\mathrm{line}} = 15^\circ$.  }
\label{fig:sgr7i8nb12_lo}
\end{figure}

\begin{figure}
\plotone{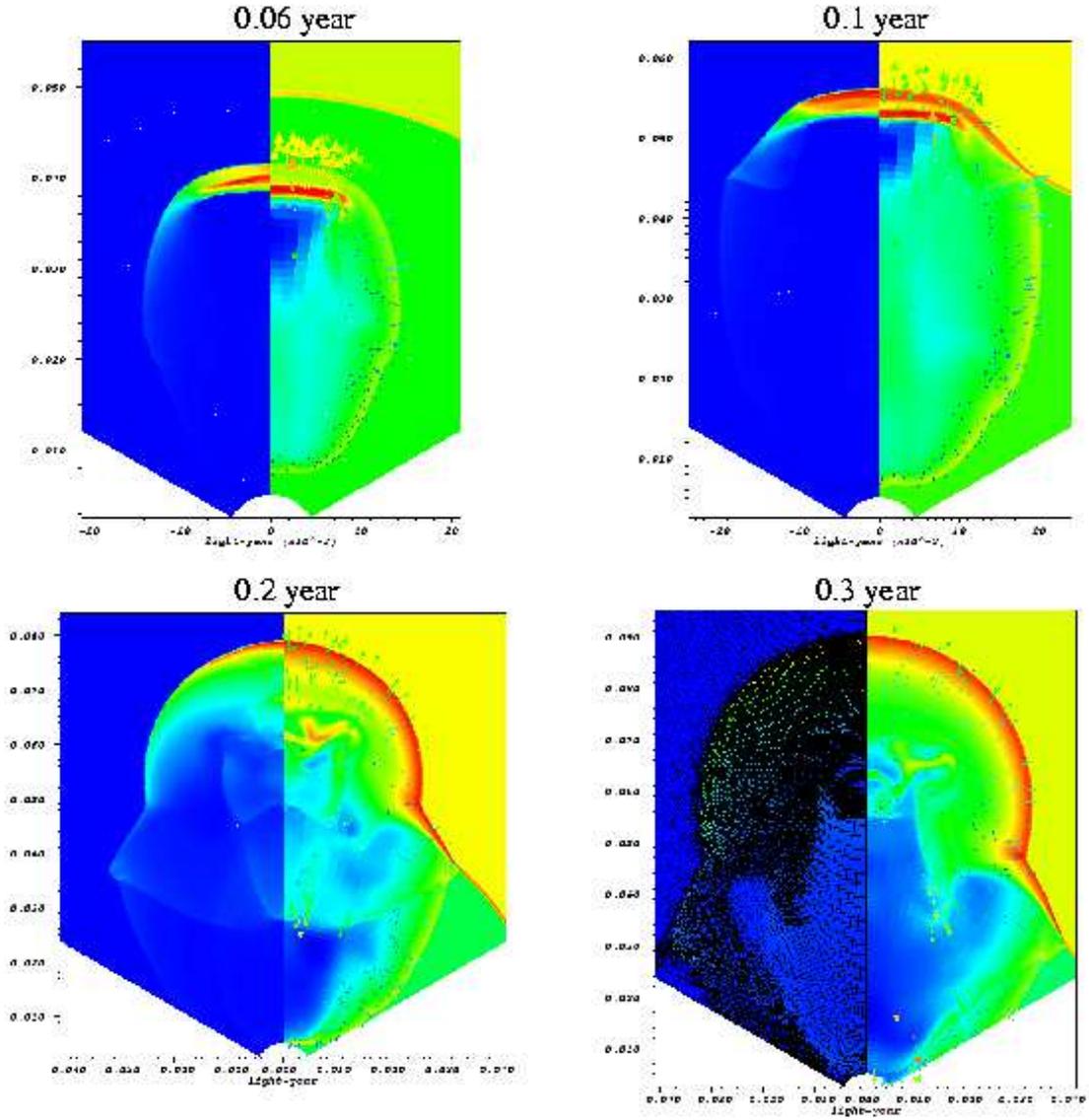} 
\caption{ Same plots as in Fig.~\ref{fig:sgr7i8nb122D}, but for the
uniform jet model J5A40U.  In contrast to Fig.~\ref{fig:sgr7i8nb122D},
the uniform ejecta is not imparted with a poleward velocity component,
and therefore does not exhibit flow focusing but instead generates a
more spherical external shock.  The left half of the final time panel
shows the adaptive mesh zoning. } \label{fig:figsgr7i8nb17ce22D}
\end{figure}

\begin{figure}
\includegraphics[width=4.5in, angle=-90]{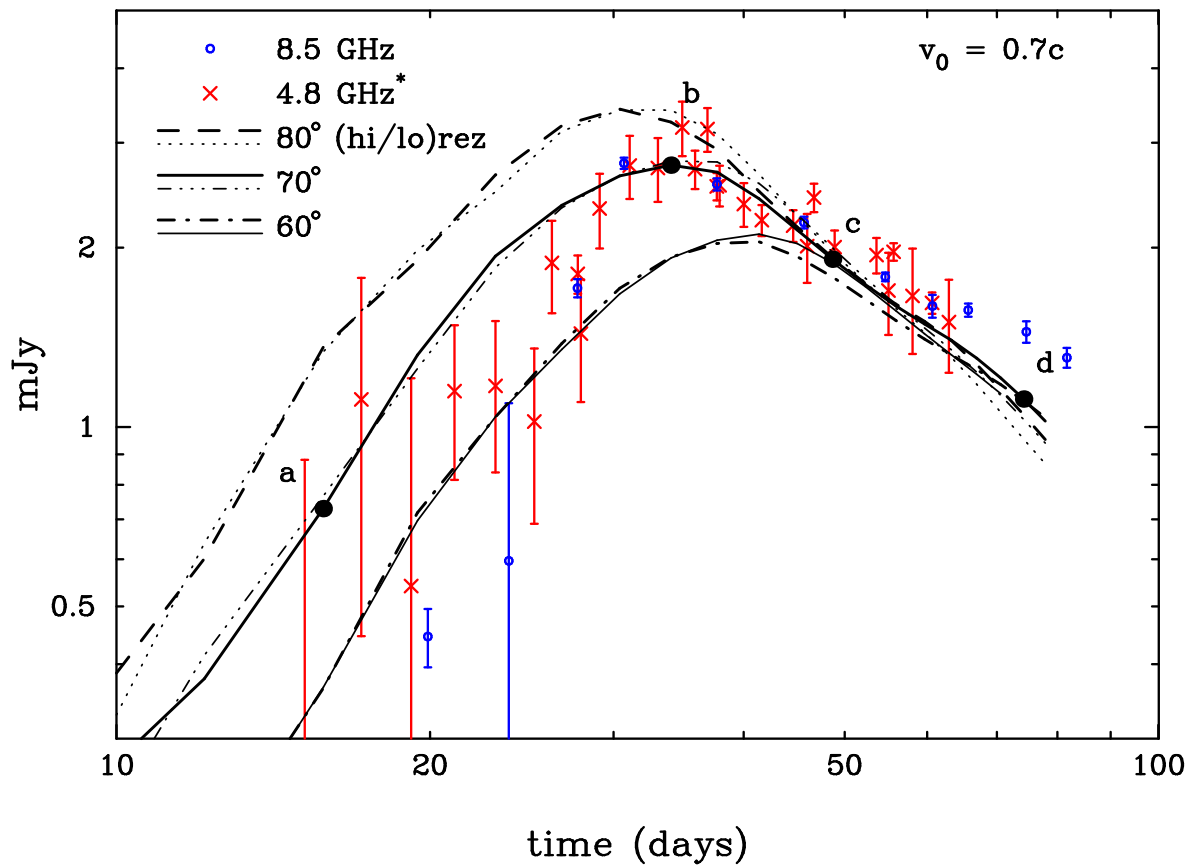} 
\caption{ Flux curves of uniform jet expansion simulations J7UHI and
J7ULO as seen from three different inclination angles, $\theta_i =
80^\circ, 70^\circ$, and $60^\circ$.  These models only differ in
their spacial resolution; J7UHI uses 5 levels of mesh refinement on
top of the base mesh, while J7ULO, after being started at high
resolution to resolve the initial motion, was deresolved to the base
mesh, and thus has zone size $2^5 = 32$ times larger than J7UHI. The
fluxes J7UHI were multiplied by 90\% (see Table \ref{tab:data}) to
best match the data, but otherwise the agreement of these two
resolutions is remarkable.  As per the discussion in
\S\ref{sec:models}, the case in which $v_0 = 0.7 c$ and $\theta_i =
70^\circ$ basically satisfies the sharp peak of the Epoch II flux
bump.  However the decay slope of these simulations tends to be
steeper than the data. The solid black dots designated {\bf a}, {\bf
b}, {\bf c}, { \bf d} are points at which flux maps are displayed in
Fig.~\ref{fig:v.7u_view}. } \label{fig:v.7u_sub}
\end{figure}

\begin{figure}
\includegraphics[width=4.5in, angle=-90]{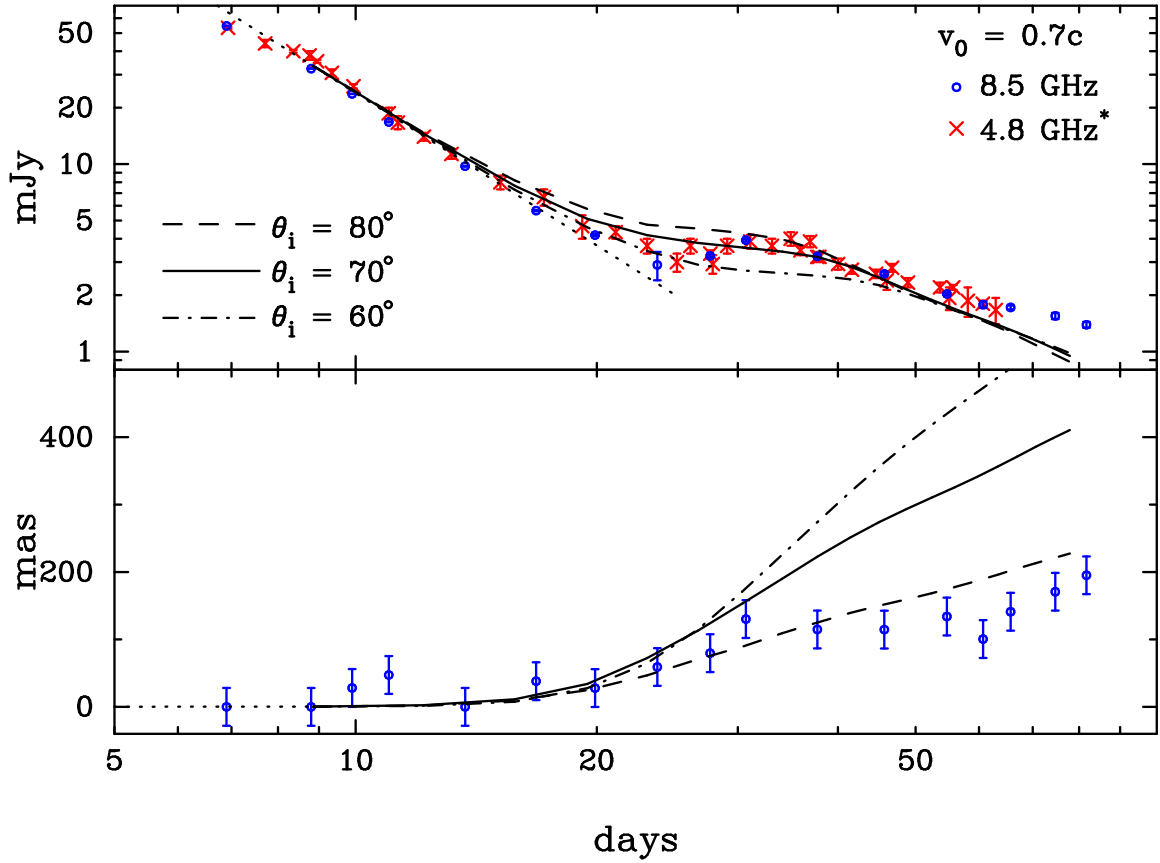} 
\caption{ Flux and centroid curves for model J7UHI seen at three
inclination angles.  While the $\theta_i = 70^\circ$ simulation
shows general agreement, note that the flux curves show a rounder,
broader peak and steeper final decay than suggested in the data.
The position of the centroid is dominated by the Epoch II component
and is quite sensitive to inclination angle. } \label{fig:v.7u_tot}
\end{figure}

\begin{figure}
\centering
\includegraphics[width=5.5in]{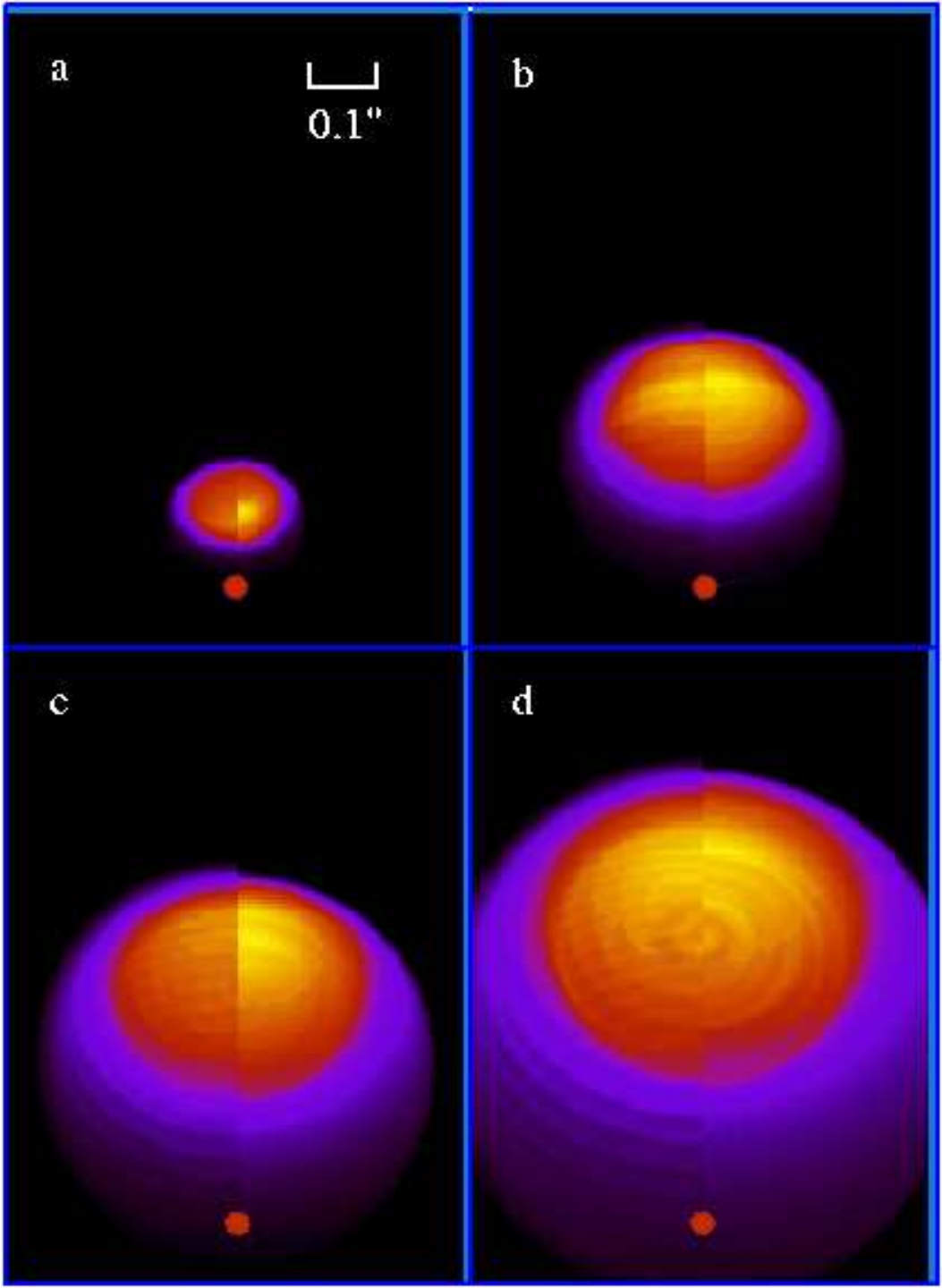} 
\caption{ Simulated flux maps of the evolution of the radio nebula at
the four times alphabetically indicated in
Fig.~\ref{fig:v.7u_sub}. Here the left half of each $0.16 \times 0.22
\mathrm{ ly}^2$ panel is from simulation J7ULO, while the right half is
from J7UHI. The dot at the lower center of each panel indicates the
position of the SGR. Both halves use the same intensity scale and
exhibit close qualitative agreement despite their large difference in
resolution. The low-resolution (left) simulation shows two sets of
discrete concentric bands centered on the jet and the SGR, indicating
its limitation in spatial and thus temporal resolution,
respectively. As shocks bounce between the ejected mass shell and the
forward shock, their transient flux is evident as broad bands
concentric on the SGR; a most obvious example of which is seen in
panel b).  }
\label{fig:v.7u_view}
\end{figure}

\begin{figure}
\plotone{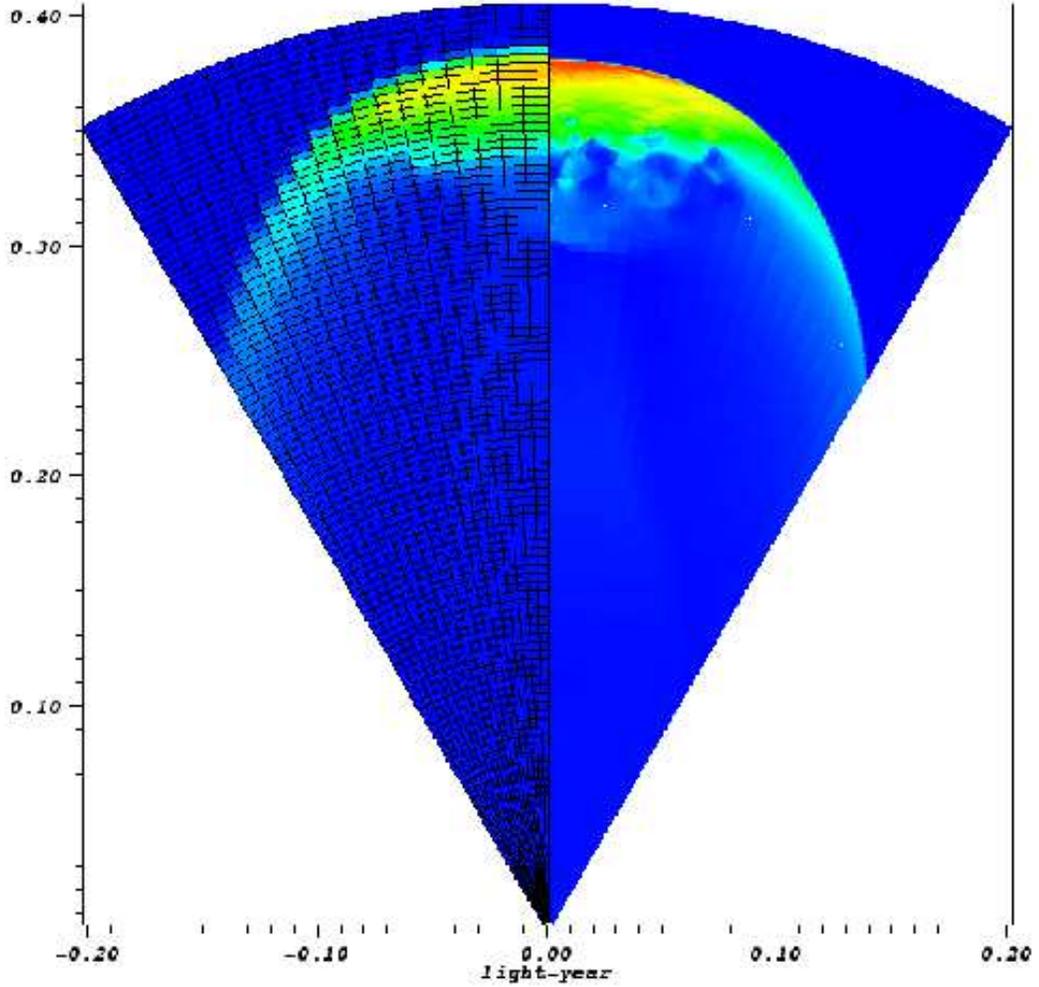} 
\caption{ A comparison of proper energy densities for simulations
J7UHI (right) and J7ULO (left) at their completion time: 0.6 year.
They are plotted on the same color scale.  One can see the external
shock has driven upward and expanded from its initial position, $r_0
= 0.01$ light-year.  Both simulations used the same base mesh, shown
on the left for reference, but J7UHI allowed 5 levels of refinement
on top of this mesh.  The structure in the flow behind the shock, at
radius $\sim 1/3$ ly, is due to Rayleigh-Taylor instabilities
disrupting the original disk of ejected mass. } \label{fig:v.7u_t.6}
\end{figure}

\begin{figure}[htb]
\centering
\includegraphics[width=4.5in]{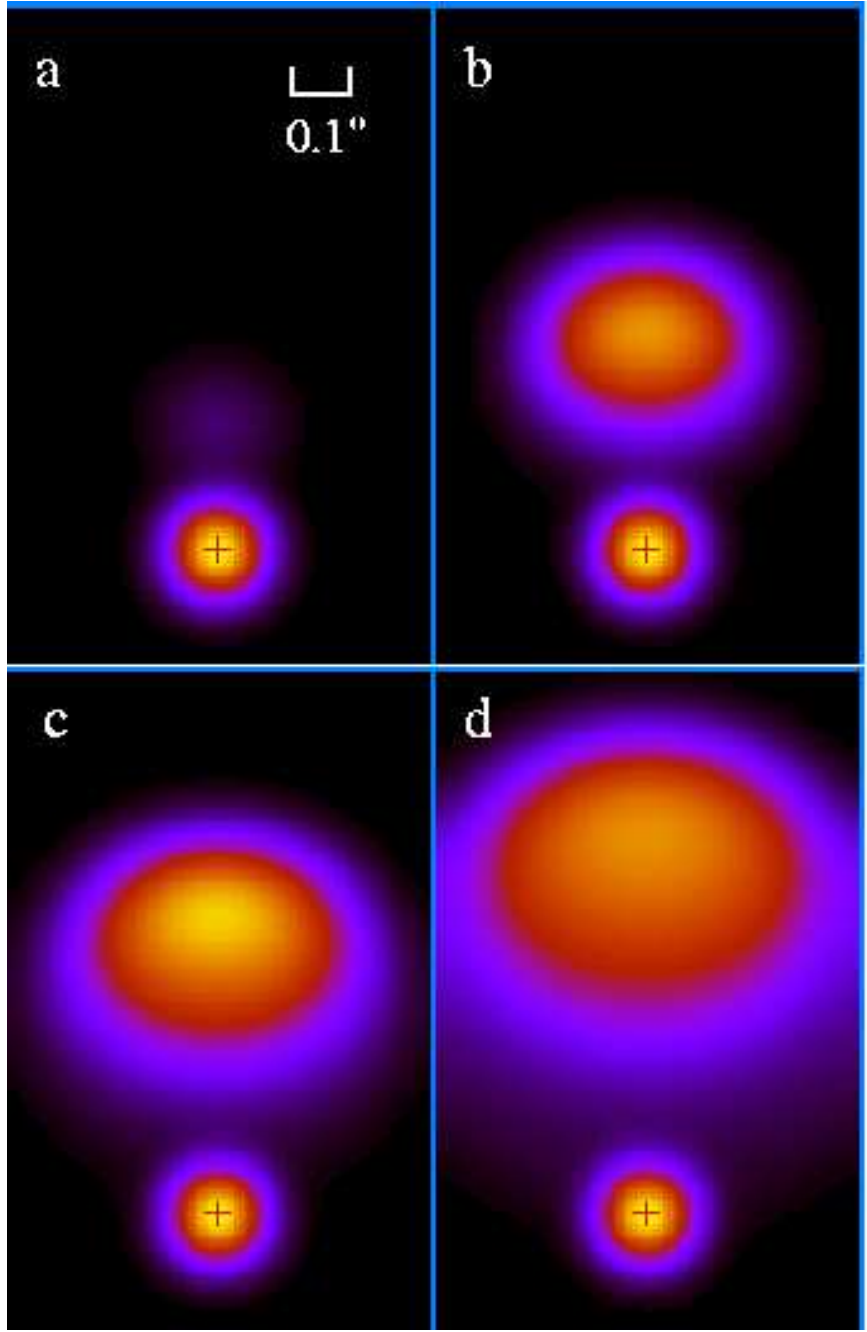} 
\caption{ Model J7UHI for the same sequence of images as in
Fig.~\ref{fig:v.7u_view}, but with the Epoch I (eqn.~\ref{E:FI})
flux component added at the position of the SGR (denoted by the red
``+'') and the image is then convolved with 50 mas gaussian to
simulate resolution limitations.  The panel dimensions, $0.16 \times
0.25 \mathrm{ ly}^2$, are slightly enlarged from those of
Fig.~\ref{fig:v.7u_view} to accommodate the blurring. }
\label{fig:v.7smear}
\end{figure}

\begin{figure}[htb]
\includegraphics[width=4.5in, angle=-90]{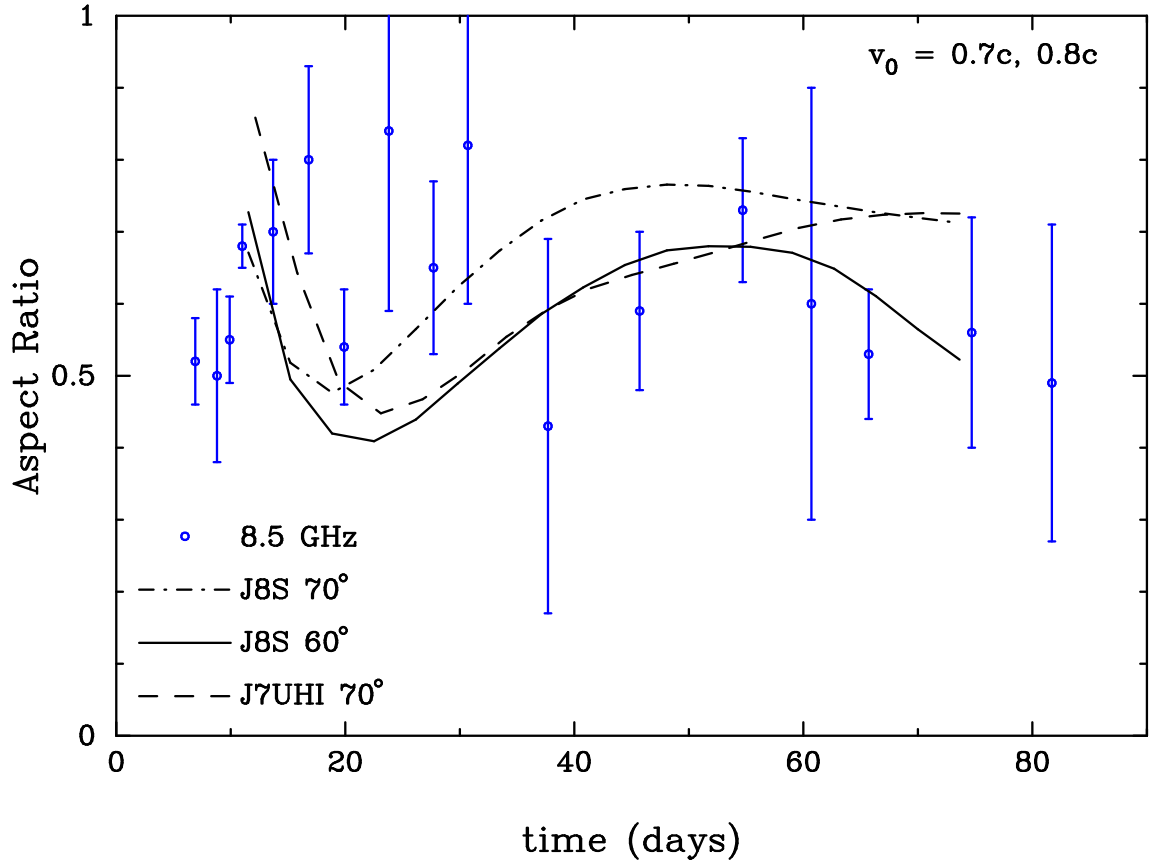} 
\caption{ Aspect ratios for higher velocity simulations show the
same qualitative behavior as those in Fig.~\ref{fig:aspectv.5}.
Because these models exhibit a more gradual brightening of Epoch II
(Fig.~\ref{fig:v.7u_sub}), the aspect ratio decreases earlier and
has a smoother inflection before increasing. } \label{fig:aspectv.7}
\end{figure}

\begin{figure}[htb]
\includegraphics[width=4.5in, angle = -90]{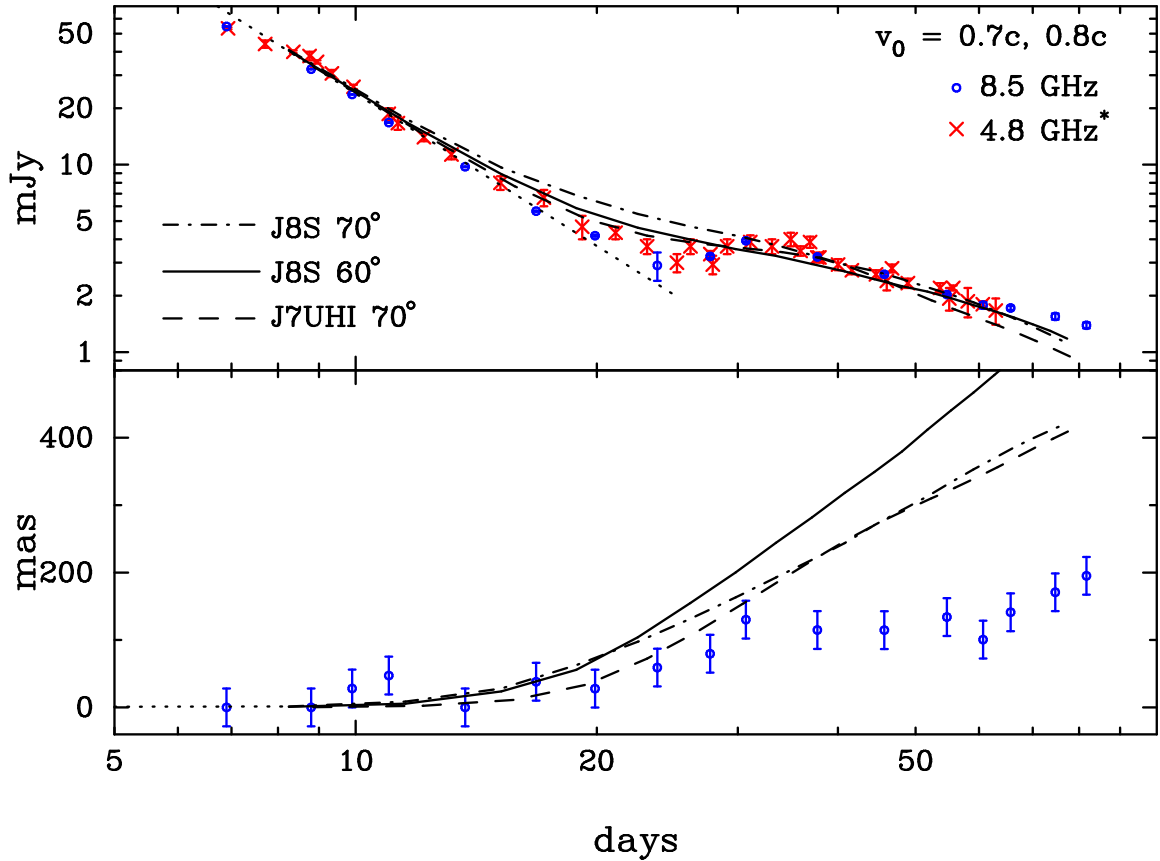} 
\caption{ Flux and centroid curves for model J8S seen at two
inclination angles $60^\circ$ and $70^\circ$.  J7UHI is shown at
$70^\circ$, as shown in Fig.~\ref{fig:v.7u_tot}.  Note that, while
none of these simulations match the data during the flux rise of
Epoch II, $\sim 20 - 30$ days, the structured model J8S gives a final
flux decay slope that is more consistent with the observed
$t^{-1.1}$. } \label{fig:v.7last_tot}
\end{figure}

We have considered two models for how the Epoch II flux bump
evolved. First, the collisional brightening model assumes that the
rapid brightening of Epoch II is caused by the ejecta colliding with
a wall or step in ISM density (\S\ref{sec:collision}).  This model
is very successful at producing the observed rapid flux brightening
which precisely coincides with a rapid centroid motion
(Fig.~\ref{fig:v.5_tot}) and rapid decrease in aspect ratio
(Fig.~\ref{fig:aspectv.5}).  Indeed, three distinct models of this
class are presented, J5A355, J5A40S and J5A40G, all of which
satisfy the data despite their varied inclination angles, jet
structures, and total energies.  Distinct by design, these models
demonstrate some of the degeneracy in the observable
characteristics. However, these models do constrain the initial
ejecta velocity to be $v_0 \sim 0.5 c$. They also suggest that if
there is a ``wall'' in the ISM, it must be several $\times 10^{16}$
cm from the SGR. If $v_0$ was much below this value, it becomes
difficult to satisfy the centroid offset data
(e.g.~Figs.~\ref{fig:v.35_tot} \&\ref{fig:v.377_tot}), but if $v_0$
was much in excess of $0.5 c$, implying a higher inclination angle
to keep the centroid data consistent, one finds the flux to brighten
more rapidly than is observed and a tendency for the late time flux
decay rate to become steeper, due to Doppler contraction, than the
$t^{-1.1}$ that is observed.  Thus we believe these three models to
trace out the locus of acceptable parameters for the collisional
brightening paradigm.

It is interesting to note that the most successful models in this
class had $n_{int}$ differing from $n_{ext}$ by only a factor of $\sim
4$.  It may well be that models in which $n_{int} \approx n_{ext}$ can
be made to work as well, therefore implying that, rather than reaching
the terminus of a void, the Epoch II brightening might have occurred
when the ejecta collided with a small cloud of material.

The second class of models we consider are Doppler brightened.  In
these models any collision with a density wall or discontinuity in the
ISM is assumed to happen prior to any observations ($< 7$ days) and so
cannot be responsible for the rapid flux brightening of Epoch II.
Therefore the brightening is only due to expansion of the shock
forward of the ejecta as it plows into the ISM.  Thus the only way to
control the shock brightening is by Doppler contraction of the
evolution as seen by an observer at high angle of inclination (\S
\ref{sec:doppler}).  Thus these models require higher $v_0$ and
$\theta_i$.  The two models presented which best match the data are
J7UHI and J8S.  Both qualitatively match the data, but miss the
finer features in Epoch II: the rapid flux brightening coupled with
rapid centroid motion and aspect ratio decrease.  To improve the fit
in these respects, the initial ejecta velocity and inclination angle
need to be higher than what was used here $v_0 > 0.8c$, $\theta_i >
70^\circ$.  Since increasing these parameters makes the final decay
slope steeper, a structured jet morphology, as in model J8S, becomes
more likely.  Finally, one could appeal to spectral evolution effects
such as a hardening spectral index $p$ (equation~\ref{eqn:p22}) to foster a more rapid brightening.  So although these Doppler brightened
models do not fit the data as well as the collision-brightened models
(with their extra degrees of freedom associated with the ISM wall),
they might be improved with added physics in future work.

A key difference between these two models is the scale of the external
density $n_{ext}$.  As can be seen in Table \ref{tab:data}, the
collisional brightening models J5A355, J5A40S and J5A40G have external
densities greater than 100 times larger than those of the Doppler
brightened models J7UHI and J8S, but all these models have a similar
total energy, $E_0 \sim 10^{46}$ ergs.  Considering the scaling of
eqn.~\ref{E:scale}, $n_{ext} \propto E_0$,  an independent
constraint on the external density might help discriminate between the
two models.  For instance \citet{grrte+06} place a lower
bound, $n_{ext} \geq 10^{-2}$, which requires $n_{ext}$ and thus
$E_0$ be scaled at least three times higher than quoted in Table
\ref{tab:data}.  On the other hand, an upper bound on $n_{ext}$ would
require $n_{ext}$ and thus $E_0$ to be scaled down for the collisional
brightening models, which requires improved shock radiation efficiency
(eqn.~\ref{eqn:enscale}) which might become physically untenable.

Finally, in all of the models we discuss here we find that a
non-uniform jet, with a velocity profile decreasing with angle from
the jet axis, is required to fit the decay slope of Epoch II,
$t^{-1.1}$.  As discussed in \S\ref{sec:structuredjetresults}, this is
the result of ``flow focusing'' behind the external shock, which
maintains the strength of this shock and thus extends the duration of
emission.  While some basic tuning was required (see Table
\ref{tab:data}) to fit the data, the general feature of flow focusing
seems very generic and ubiquitous for non-uniform velocity
distributions.  To further explore and test this idea, it would be
useful to have late time ($> 80$ days) flux measurements to catenate
onto those used here and compare with similarly extended
simulations.  It is very reasonable to imagine that jetted outflows
generated by nature will be faster at the core than in the wings.
Thus this effect might appear in other jetted phenomena such as
gamma-ray bursts or jetted outflows from active galactic nuclei.  It
will be the subject of future work to explore other instances of this
mechanism.

\begin{acknowledgements}
This work was performed under the auspices of the U.S. Department of
Energy by University of California Lawrence Livermore National
Laboratory under contract W-7405-ENG-48. PCF gratefully acknowledges
the support of a Faculty R\&D grant from the College of Charleston.
\end{acknowledgements}

\clearpage
\bibliographystyle{apj}


\clearpage

\begin{deluxetable}{lrrrrrrrrcccr}
\tabletypesize{\footnotesize} \tablecaption{Table of simulations. \label{tbl-1}}
\tablewidth{0pt} \tablehead{ \colhead{Name} & \colhead{$v_0/c$} &
\colhead{$\theta_0$} & \colhead{jet (s)\tablenotemark{(a)}} &
\colhead{$M_0$\tablenotemark{(b)}} & \colhead{$E_0$
\tablenotemark{(c)}} & \colhead{$\mathcal{E}_N$\tablenotemark{(d)}}
& \colhead{$r_0$\tablenotemark{(e)}} &
\colhead{$r_{wall}$\tablenotemark{(e)}} &
\colhead{$n_{ext}$\tablenotemark{(f)}} &
\colhead{$n_{int}$\tablenotemark{(f)}} &
\colhead{$n_{wall}$\tablenotemark{(f)}} & \colhead{Figures} }
\startdata
\cutinhead{Collisional Brightening Model}
J35A    & 0.35  & $24^\circ$ & P & 3.56 &3.26 & 1.15 & 2 & 3.15 & 1.0 & $10^{-6}$ & 0.6 & \ref{fig:v.35_sub}-\ref{fig:v.35_tot} \\
J377I25 & 0.377 & $24^\circ$ & P & 3.08 & 2.77 & 1.05 & 2 & 3.15 & 1.0 & 0.26      & 6.0 & \ref{fig:v.377_tot} \\
J377I30 & 0.377 & $24^\circ$ & P & 3.08 & 2.77 & 1.05 & 2 & 3.15 & 1.0 & $10^{-6}$ & 6.0 & \ref{fig:v.377_tot} \\
J377I35 & 0.377 & $24^\circ$ & P & 3.08 & 2.77 & 1.15 & 2 & 3.15 & 1.0 & $10^{-6}$ & 0.6 & \ref{fig:v.377_tot} \\
J5A355  & 0.5   & $24^\circ$ & P & 2.53 & 2.35 & 0.7  & 2 & 3.5  & 1.04& 0.26      & 5.1   & \ref{fig:v.5_tot},\ref{fig:plotA}-\ref{fig:sgr7i8nb12_lo} \\
J5A40S  & 0.5   & $24^\circ$ & P (0.5) & 1.56 & 1.5  & 0.95 & 1 & 4.9  & 0.53& 0.1       & 1.25   & \ref{fig:v.5_tot}-\ref{fig:aspectv.5} \\
J5A40U  & 0.5   & $12^\circ$ & U & 0.2 & 0.21  & 0.95 & 1 & 4.9  & 0.53& 0.1       & 1.25   & \ref{fig:v.5_sub}-\ref{fig:aspectv.5},\ref{fig:figsgr7i8nb17ce22D} \\
J5A40G  & 0.5   & $15^\circ$ & G (0.75) & 1.09 & 1.03 & 1.15 & 1 & 4.9  & 0.83& 0.21       & 1.3   & \ref{fig:v.5_tot},\ref{fig:plotA}-\ref{fig:aspectv.5} \\
J5A40SE2& 0.5   & $24^\circ$ & P (0.5) & 3.13 & 3.0  & 0.4  & 1 & 4.9  & 0.53& 0.1       & 1.25   & \ref{fig:v.5_sub} \\
J5A40SE.5& 0.5  & $24^\circ$ & P (0.5) & 0.78 & 0.75  & 2.7  & 1 & 4.9  & 0.53& 0.1       & 1.25   & \ref{fig:v.5_sub} \\
J5A40SEN2& 0.5  & $24^\circ$ & P (0.5) & 3.13 & 3.0  & 0.271  & 1 & 4.9  & 1.06& 0.2       & 2.5   & \ref{fig:v.5_sub} \\
\cutinhead{Doppler Brightening Model}
J7UHI   & 0.7   & $12^\circ$ & U & 0.6   & 0.76   & 0.9  & 1 & 1.1  & 0.003 & $3 \times 10^{-7}$ & 0.003   & \ref{fig:v.7u_sub}-\ref{fig:v.7last_tot}\\
J7ULO   & 0.7   & $12^\circ$ & U & 0.6   & 0.76   & 1.0  & 1 & 1.1  & 0.003 & $3 \times 10^{-7}$ & 0.003   & \ref{fig:v.7u_sub},\ref{fig:v.7u_view}-\ref{fig:v.7u_t.6} \\
J8S     & 0.8   & $20^\circ$  & P (0.25) & 1.5  & 1.9 & 0.8  & 1 & 1.1  & 0.003 & $3 \times 10^{-7}$ & 0.003   & \ref{fig:aspectv.7}-\ref{fig:v.7last_tot}\\
\enddata

\tablenotetext{(a)}{Jet type: (U)niform, (P)ower-law, (G)aussian and
velocity power, $s$ (=1 unless specified in parenthesis). See
\S\ref{sec:structjets}. } \tablenotetext{(b)}{$\times 10^{25}$ gm }
\tablenotetext{(c)}{$\times 10^{46}$ ergs } \tablenotetext{(d)}{Fine
tuning parameter to match simulated flux to data (see
\S\ref{sec:compare}). Multiply $M_0, E_0, n_{ext}, n_{int},
n_{wall}$ by $\mathcal{E}_N^{1/1.8}$. } \tablenotetext{(e)}{$\times
10^{-2}$ light-year} \tablenotetext{(f)}{baryons cm$^{-3}$}
\label{tab:data}
\end{deluxetable}
\end{document}